\documentclass[12pt]{article}
\usepackage{authblk}
\usepackage{geometry}
\usepackage{amsmath}
\usepackage{amssymb}
\usepackage{amsthm}
\usepackage{mathrsfs}
\usepackage{subfigure}
\usepackage{customcommands}
\usepackage{bm}
\usepackage{Nettastyle}
\usepackage{dsfont}

\usepackage{amsmath}
\usepackage{amssymb}
\usepackage{amsfonts}
\usepackage{braket}
\usepackage[normalem]{ulem}
\usepackage{color}
\usepackage{bbold}
\usepackage{ulem}
\usepackage{mathtools}
\usepackage{tensor} 

\preprint{MIT-CTP/5304}

\title{Finding Pythons in Unexpected Places}

\author[1]{Netta Engelhardt,}
\author[2, 3]{Geoff Penington,}
\author[4]{and Arvin Shahbazi-Moghaddam}

\affiliation[1]{Center for Theoretical Physics, Massachusetts Institute of Technology, \\Cambridge, MA 02139, USA}
\affiliation[2]{Center for Theoretical Physics and Department of Physics,\\
University of California, Berkeley, CA 94720, U.S.A. and}
\affiliation[3]{Institute for Advanced Study, 1 Einstein Dr, Princeton, NJ 08540, U.S.A.} 
\affiliation[4]{Stanford Institute for Theoretical Physics,\\ Stanford University, Stanford, CA 94305 USA}
\emailAdd{engeln@mit.edu}
\emailAdd{geoffp@berkeley.edu}
\emailAdd{arvinshm@gmail.com}

\abstract{We argue that novel (highly nonclassical) quantum extremal surfaces play a crucial role in reconstructing the black hole interior even for isolated, single-sided, non-evaporating black holes (i.e. with no auxiliary reservoir). Specifically, any code subspace where interior outgoing modes can be excited will have a quantum extremal surface in its maximally mixed state. We argue that as a result, reconstruction of interior outgoing modes is always exponentially complex. Our construction provides evidence in favor of a strong Python's lunch proposal: that nonminimal quantum extremal surfaces are the exclusive source of exponential complexity in the holographic dictionary. We also comment on the relevance of these quantum extremal surfaces to the geometrization of state dependence in the typicality arguments for firewalls.}

\begin{document}
\maketitle

\section{Introduction} \label{sec:intro}

The recent renaissance in the black hole information frontier, starting with~\cite{Pen19, AEMM}, has unveiled a new understanding of the geometrization of unitarity and computational complexity of the Hawking radiation. These developments were catalyzed by the discovery~\cite{Pen19, AEMM} and subsequent justification~\cite{PenShe19, AlmHar19} of the existence of novel quantum extremal surfaces (QESs)~\cite{EngWal14}: QESs that are nonperturbatively distinct from their classical counterparts.

Recall that a QES is a surface $\gamma$ that extremizes the generalized entropy~\cite{EngWal14}:
\begin{align}
    S_\text{gen}[\gamma] = \frac{A}{4G_N} + S,
\end{align}
under local deformations of $\gamma$. Here A is the area of $\gamma$ and $S$ is the entropy of quantum fields outside of $\gamma$.

The novel QES phenomenon in the semiclassical regime is a consequence of large entropy gradients that compete with leading variations in the area term.  The existence of such QESs is directly responsible for the turnover in the unitary Page curve~\cite{Pag93b}, which has since led to new questions about the emergence of the semiclassical description of gravity and the contribution of nonperturbative effects to the gravitational path integral (see work starting with~\cite{PenShe19, AlmHar19}). These new nonclassical QESs also show that the interior of the black hole can be reconstructed from its Hawking radiation after the Page time~\cite{Pen19, AEMM, AlmMah19a} and that this reconstruction is exponentially complex when executed by an outside observer~\cite{BroGha19}. The existence of novel QESs therefore has the potential to significantly elucidate upon one of the main challenges in the holographic dictionary: reconciling how seemingly simple and natural quantities in the bulk, such as local bulk operators, can be very complicated: in fact, they can be exponentially complicated -- in the number of boundary degrees of freedom (or $1/G_N$) -- in the boundary description.

\begin{figure}
    \centering
    \includegraphics[width=0.45\textwidth]{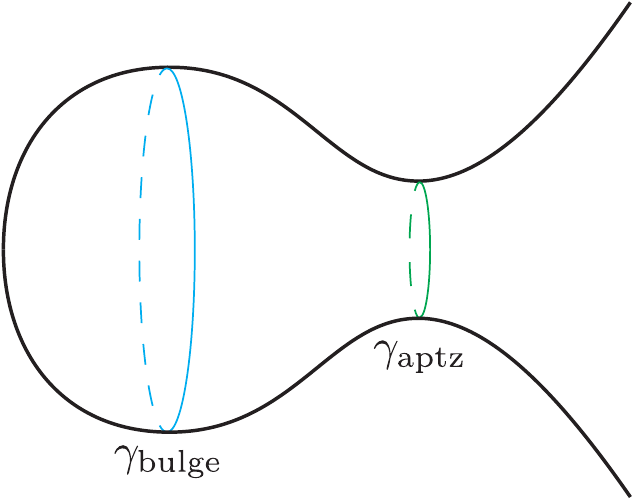}
    \caption{A time-symmetric slice of a one-sided Python's lunch geometry involving two quantum extremal surfaces $\gamma_{\text{aptz}}$ (or the outermost quantum extremal surface) and $\gamma_{\text{bulge}}$. The slice asymptotes to the boundary of AdS on the right. Since the quantum minimal extremal surface is empty, the entire geometry is in the entanglement wedge of the boundary CFT, but the region behind $\gamma_{\text{aptz}}$ is encoded in it with exponential complexity. The exponent of the complexity is given by half of the difference between the generalized entropies of $\gamma_{\text{bulge}}$ and $\gamma_{\text{aptz}}$.}
    \label{fig-onelunch}
\end{figure}

Concretely, it was proposed in~\cite{BroGha19} that certain examples of such extraordinary complexity could be explained by the existence of nonminimal QESs in the geometrical configuration of a ``Python's lunch'', or bulge, in the bulk spacetime geometry; see Fig.~\ref{fig-onelunch} for an example. In tensor network toy models of AdS/CFT, there is strong evidence that a Python's lunch geometry implies exponential reconstruction complexity due to postselection. The conjecture of~\cite{BroGha19}, backed up by a number of examples, was that this is also true in gravity.

The results in this paper provide significant evidence in favor of a stronger position: that nonminimal QESs are in fact the source of \emph{all} exponential complexity in the holographic dictionary. \footnote{Here we mean exponential in natural bulk geometric quantities. Thus the long-wormhole constructions in which the volume of the wormhole is exponentially large in $G_{N}^{-1}$ do not constitute counterexamples to this proposal because the complexity is still linear in the volume of the wormhole.}

In recent work~\cite{EngPen21}, we gave a very general argument that, in the limit where the bulk physics can be treated classically, the ``no-man's land'' between the outermost extremal surface and the event horizon is always simply reconstructible. Put differently, there exists a simple algorithm that converts a state in which the region between the outermost classical extremal surface and the event horizon is nonempty into a state in which the classical extremal surface lies on (or limits to) the event horizon. We therefore concluded that reconstruction in classical bulk spacetimes is indeed easy so long as the bulk operator is not inside a Python's lunch.\footnote{To be clear, our results in~\cite{EngPen21} only show that reconstruction of operators outside a lunch is simple while reconstructing operators inside it is not. We did not give a quantitative estimate of exactly how difficult it is to reconstruct operators inside a lunch, which was an important part of the original conjecture of~\cite{BroGha19}.} 

To support such a statement in semiclassical gravity, however, we must contend with the challenge that until now, examples of novel QESs in AdS have been restricted to black holes coupled to a reservoir. This is needed to make the black hole evaporate: isolated AdS black holes do not evaporate (unless they are parametrically small) due to Hawking radiation being reflected back into the black hole. One might therefore wonder whether these nonclassical surfaces are solely artifacts of coupling the AdS bulk to an auxiliary system, or alternatively are a very specific consequence of black hole evaporation, and do not play a role in black hole physics otherwise. Either possibility would call into question the generality of insights derived about the AdS/CFT dictionary from novel QESs, and in particular would challenge the validity of the strong Python's lunch proposal. 

\begin{figure}
    \centering
    \includegraphics[width=0.4\textwidth]{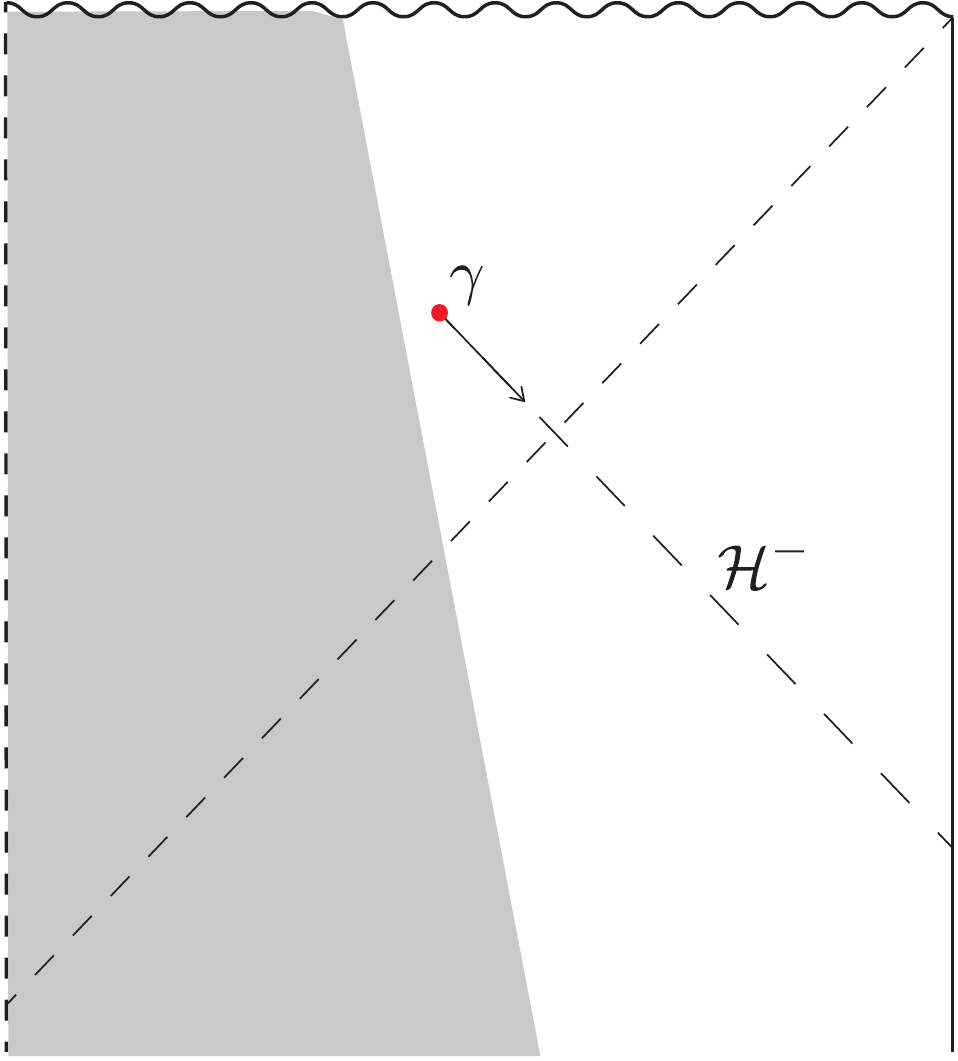}
    \caption{A quantum extremal surface $\gamma$ in the future of the asymptotic boundary in an isolated black hole is forbidden by the generalized second law. When we have spherical symmetry, the quantum extremal surface $\gamma$ would lie on a past horizon $\mathcal{H}^-$, which by the generalized second law must generically have positive quantum expansion towards the boundary -- in contradiction with the vanishing quantum expansion which would follow from the quantum extremality of $\gamma$. The non-spherically-symmetric argument works analogously~\cite{EngWal14}.}
    \label{fig-causal}
\end{figure}

In fact, isolated non-evaporating black holes (with no reservoir) prima facie appear to be an obvious counterexample to our proposal. For black holes that have been allowed to equilibrate for more than the scrambling time, various arguments -- for instance the transplanckian precursor problem (originally discussed in~\cite{Unr76} in general and in~\cite{HeeMar12} in AdS/CFT in particular), the quantum mechanics of fast scrambling systems~\cite{HayPre07, SekSus08} etc. -- suggest that the reconstruction of interior outgoing modes should be exponentially complex. However, non-evaporating black holes formed from collapse are \emph{forbidden}~\cite{EngWal14} from having nontrivial QESs by the Generalized Second Law~\cite{Bek72, Wal11}, since the entire spacetime lies in the causal future of the asymptotic boundary (see Fig.~\ref{fig-causal}). The apparent conclusion is that bulk reconstruction can be exponentially complex if the reconstructed operator is behind the event horizon, even if it is not behind a quantum extremal surface.

Here we show that this conclusion is too fast. Even though the post-collapse black hole state itself has no nontrivial QES (and hence no Python's lunch), we cannot talk about the complexity of reconstructing interior outgoing modes until we introduce a code subspace where those modes can be excited. Once we do so, we are forced to consider states (in particular the maximally mixed state within the code subspace) where outgoing modes are disentangled across the horizon. The primary technical aspect of this paper will be to show that this  disentanglement creates an entropy gradient, which in turn nucleates a novel QES. The novel QES is the appetizer of a ``secret'' Python's lunch -- invisible until you consider reconstructing an interior outgoing mode -- that explains the exponential reconstruction complexity. 

We emphasize that this certainly does not constitute an extension of our classical \textit{proof} of the strong Python's lunch conjecture to arbitrary semiclassical spacetimes. That would be a very difficult task, although see \cite{Levine:2020upy} for an attempt in the special case where the ``no-man's land'' in the original spacetime is perturbatively small. 

However, our argument does show that a naive objection to our strong Python's Lunch proposal on the basis of the absence of nontrivial quantum extremal surfaces in single-sided non-evaporating black holes (and other similar examples) is unfounded. More generally the existence and importance of these hidden lunches shows that the highly nonclassical QESs that have already been observed in evaporating black holes~\cite{Pen19, AEMM} are part of a much broader and more generic phenomenon involving black holes in semiclassical gravity, and are certainly not just an artifact of coupling to the auxiliary reservoir. 

We now explain our construction, focusing in particular on the role of the maximally mixed state, and then outline the structure of the paper. 

\subsubsection*{Finding the Hidden Python} 
Let us briefly review the idea of entanglement wedge reconstruction~\cite{CzeKar12, Wal12, FauLew13, HeaHub14, JafLew15, DonHar16, Har16,  FauLew17, CotHay17} in somewhat more depth as it is critical for understanding why the maximally mixed state is of relevance in the computation of the complexity. Bulk reconstruction is most naturally understood within the framework of quantum error correction~\cite{AlmDon16, Har16, HayPen18, AkeLei19}. Specifically, the action of a bulk operator is in general only defined for a code subspace of states with a particular semiclassical bulk geometry. The embedding of this bulk code subspace within the larger boundary CFT Hilbert space means that there can exist many distinct CFT operators that `reconstruct' a given bulk operator (i.e. act correctly on states in the code subspace). 

Often we are interested in whether a bulk operator can be reconstructed by a boundary operator that acts only within a particular subregion of the boundary. Roughly speaking, reconstruction is possible when the bulk operator in question lies within the entanglement wedge -- the region between the minimal QES and the boundary -- of that boundary subregion. 

Crucially, however, as emphasized in~\cite{HayPen18, AkeLei19}, the reconstructibility of a bulk operator depends on the code subspace for which we want the reconstruction to work. Explicitly, the bulk operator needs to lie within the entanglement wedge for the maximally mixed state on that code subspace.\footnote{Technically, depending on whether one uses average or worst-case error to define the reconstruction accuracy, the correct condition is either that the operator lies within the entanglement wedge for the maximally mixed state, or that it lies within the entanglement wedge for all states (pure or mixed) within the code subspace. For our purposes (working in a single semiclassical background and studying states where the naive QES prescription (see~\cite{AkePen20} for deviations from regime) is valid), these two definitions are effectively equivalent.} Since the generalized entropy, and hence the minimal QES, depends on this choice of bulk state through the bulk entropy term, each choice of code subspace can give a different entanglement wedge, and hence a different set of reconstructible bulk operators.

The Python's Lunch conjecture, which is motivated by tensor network toy models, works in a very similar way. As we explain in Sec. \ref{sec:PLreview}, the conjecture suggests that the complexity of reconstructing a bulk operator, for a particular choice of code subspace, depends on whether the operator lies behind a nonminimal quantum extremal surface -- i.e. inside a lunch. In particular, by doing a careful reanalysis of the original arguments for the conjecture, we show that the relevant question is whether the operator lies behind a nonmiminal QES \textit{for the maximally mixed state on the chosen code subspace}. If it does, the reconstruction complexity is exponential in the ``size'' of the lunch; i.e.  the difference between the generalized entropies of two quantum extremal surfaces -- the ``bulge'' surface and the ``appetizer'', or outermost, surface. See Fig.~\ref{fig-onelunch}. Again, we show that these generalized entropies need to be evaluated in the maximally mixed state. To complete the section, we extend the Python's Lunch conjecture to predict the complexity of reconstruction in the presence of multiple lunches.

In Sec. \ref{sec:maincalc}, we construct a code subspace of an arbitrary (non-evaporating) black hole formed from collapse and not coupled to a reservoir; such black holes are expected though not proven to approach Kerr-AdS in the late-time adiabatic regime (see~\cite{DafRod13} for a review). We thus assume its geometry at the horizon at late times is approximately Schwarzschild-AdS for simplicity (and we anticipate our results generalize to Kerr-Neumann). Zooming in on a neighborhood of the event horizon, we build our code subspace from outgoing Rindler-like wave-packets~(i.e. Hawking wave-packets).  Our goal is to demonstrate the existence of novel nonminimal QESs that nucleate for the maximally mixed state (and the thermal state) within this code subspace. 

Because the Rindler modes are disentangled, this mixed state has a larger bulk entropy gradient than the Hartle-Hawking state under inwards deformations of a bulk entangling surface near the horizon. Moreover, as we move the surface backwards in time along the horizon, this entropy gradient is blueshifted, and ends up dominating over the classical contribution to the generalized entropy variation, i.e. the classical expansion times $1/4G_N$. The end result is that surfaces slightly behind the horizon and more than a scrambling time in the past have negative quantum expansion under any outward deformation. We then invoke a result from~\cite{BroGha19, BouSha21} (based on the ``maximin'' construction~\cite{Wal12, MarWal19, AkeEng19b}) that such surfaces contain a quantum extremal surface in their exterior. See Fig.~\ref{fig:disentangle}. This is the ``hidden'' QES that we set out to find. The associated Python's lunch provides a geometrical explanation for the exponential complexity of decoding the interior modes.

In Sec. \ref{sec:toymodels}, we compare the gravity calculations from Sec. \ref{sec:maincalc} with an explicit algorithm for decoding ``interior partners'' (i.e. the purification of Hawking modes) in a quantum circuit toy model of a black hole. We find a precise quantitative agreement between the two. We also explain how this toy model relates to the tensor network discussion from Sec. \ref{sec:PLreview}.

In Sec. \ref{sec:BFV}, we discuss another example of exponential complexity arising in non-evaporating black holes. This construction, due to Bouland, Fefferman, Vazirani (BFV) argues that the set of CFT states formed by simple perturbations to the time evolution of (non-evaporating, isolated) black holes should be pseudorandom -- i.e. impossible to distinguish in subexponential time~\cite{BouFef19}. We argue that similar secret lunches arise  in this setting and that they explain the exponential complexity. Specifically, while none of the individual states in the BFV ensemble has a nontrivial QES, the entire ensemble, viewed as a single density matrix, does have a lunch. This mixed density matrix plays the role of the maximally mixed state in the code subspace, revealing the hitherto hidden lunch.

In Sec. \ref{sec:disc}, we conclude the paper by discussing various potential open questions and applications of these results. In particular, we discuss the relevance of our technical results to the firewall typicality arguments.

\section{Exponential Complexity from the Python's Lunch}\label{sec:PLreview}

In Sec.~\ref{sec:PLreviewSpecific}, we review the Python's lunch proposal and its grounding in tensor networks and also explain the importance of the maximally mixed state  in determining the reconstruction complexity. In Sec.~\ref{sec:multiple}, we introduce a slight generalization of the original Python's lunch conjecture that includes multiple bulges: breakfast, lunch, and dinner. 

\subsection{Review: the Python's Lunch} \label{sec:PLreviewSpecific}

The proposed relation between exponential complexity and nonminimal QESs can be motivated from the apparent contradiction between prior holographic complexity proposals (such as CV~\cite{StaSus14, Sus14a} or CA~\cite{BroRob15}) and the Harlow-Hayden conjecture~\cite{HarHay13} that reconstruction of the Hawking radiation is exponentially complex. As applied to the evaporating black hole, the two prescriptions differ dramatically. The resolution proposed by~\cite{BroGha19} is that the extant holographic complexity proposals compute the complexity of preparing the state using a tensor network, whereas in Harlow-Hayden observers are restricted to applying (a) unitary operators to (b) a subsystem of the degrees of freedom (specifically the Hawking radiation). The Python's lunch conjecture is a refinement of the traditional complexity proposals: it gives the unitary circuit complexity of reconstructing operators, using either the global boundary or a subsystem of it. Crucially, the conjecture takes into account the difficulty of replacing nonunitary postselection by a unitary circuit. 

The intuition for the nonminimal QESs as a fix is derived from tensor networks. In tensor network toy models of bulk reconstruction (see e.g.~\cite{HaPPY}), the network from bulk to boundary must be an (approximate) isometry; typically the individual tensors have also been taken to be isometries -- i.e. unitaries under the addition of ancilla qubits in the $|0\rangle$ state. In these setups, it is simple to push through to the boundary via a unitary. The unitary circuit complexity of bulk reconstruction implementing bulk reconstruction appears to grow linearly with the number of tensors in the networks  -- so long as we assume that each individual tensor is itself simple.

\begin{figure}
    \centering
    \includegraphics[width=0.7\textwidth]{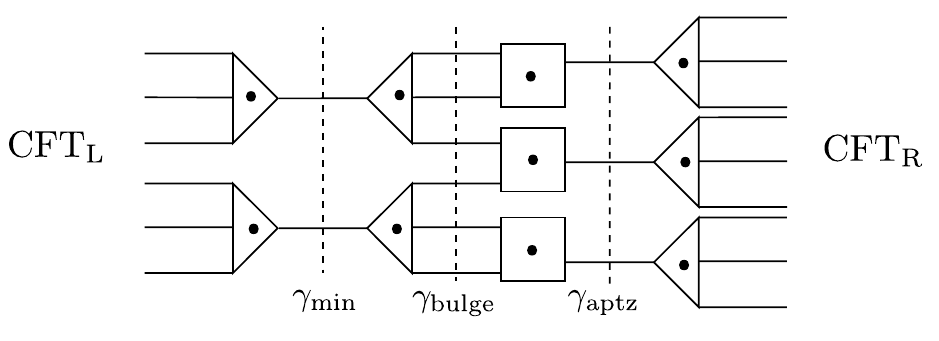}
    \caption{The Python had fish for lunch. The tensor network prepares a boundary state on left and right CFTs. The fish (triangles) are isometries while the squares involve postselection on one of the legs and the out-of-plane legs (shown with dots) represent bulk degrees of freedom. The network in particular generates an (approximate) isometry from the $\gamma_{\text{min}}$ cut, together with the bulk legs to its right, into the right CFT. The bulk legs between $\gamma_{\text{min}}$ and $\gamma_{\text{aptz}}$ are expected to be encoded on the CFT with exponential complexity. The conjectured exponent is given by half of the difference between the total bond dimensions cut through by $\gamma_{\text{bulge}}$ and $\gamma_{\text{aptz}}$, plus the bulk legs in between.}
    \label{fig:TN}
\end{figure}

This reasoning is not applicable to a tensor network in which there are individual tensors that implement postselection. Consider a tensor network consisting of a geometry with a `bulge' in the middle as shown in Fig.~\ref{fig:TN}. Starting from the left boundary, the cross section of the network first contracts to the minimal cut $\gamma_{\mathrm{min}}$, then expands within the bulge to a maximum at $\gamma_{\mathrm{bulge}}$, before contracting again to a locally, but not globally, minimal cut $\gamma_{\mathrm{aptz}}$, and finally expanding out to the right boundary. The network also includes 'bulk legs' on each tensor that represent local bulk quantum fields; the entire network forms an isometry from these bulk legs to the legs on the left and right boundary.

The tensor network shown is a toy model of a two-sided wormhole with a Python's lunch geometry. The `bulk-to-boundary' isometry described above describes the embedding of the `code subspace' of semiclassical bulk states with the correct wormhole geometry into the CFT Hilbert space(s). In the special case where $\gamma_{\mathrm{min}}$ has zero size (and we have no left boundary) then the tensor network becomes a toy model of a one-sided black hole with a Python's lunch in its interior. It is in fact this latter case that we will be of most interest to us, but it is helpful to keep the discussion more general for the moment.

Because the bulge lies to the right of the minimal cut $\gamma_{\mathrm{min}}$, bulk operators within the bulge lie in the entanglement wedge of the right boundary and should thus be reconstructible on the right boundary. This is because, for sufficiently generic tensor networks (which are the most analogous to gravity), the entire map from the minimal cut, plus bulk legs to its right, to the right boundary will be an approximate isometry, which we denote by $V$. Here we assume that the size of the locally minimal cut $\gamma_{\mathrm{aptz}}$ is larger than the size of the minimal $\gamma_{\mathrm{min}}$, plus all of the bulk legs in between the two.\footnote{By the size of a tensor network cut, we mean the number of legs, with each leg weighted by the logarithm of its dimension. For explanatory convenience, we will generally assume that all legs (both in plane legs and bulk legs) have the same dimension, so that all that matters is the number of legs. } 

Nonetheless, the tensor $P$ describing the `constriction' from the bulge $\gamma_{\mathrm{bulge}}$ (plus bulk legs between $\gamma_{\mathrm{bulge}}$ and $\gamma_{\mathrm{aptz}}$) to $\gamma_{\mathrm{aptz}}$ cannot be an isometry, because the Hilbert space dimension of the output is much smaller than that of the input. Instead, it is  the adjoint of an isometry $P^{\dagger}$ (namely the opposite-direction right-to-left map) and so can be rewritten as the combination of a unitary matrix and \emph{postselection} -- i.e. collapsing the wavefunction onto the component where certain qubits are in the state $\ket{0}$.

Since postselection is an inherently non-unitary process, we cannot use unitary gates to implement the network tensor by tensor. Instead we need to do something cleverer. The trick is to use a unitary algorithm known as Grover search, or amplitude amplification, to search for the part of the wavefunction where the qubits that need to be postselected are already in the state $\ket{0}$. 

Since Grover search is central to the Python's lunch story, it is worth reviewing how it works. Suppose we start with an initial (unknown) state $\ket{\psi}$ want to produce the state $$V \ket{\psi} = \sqrt{A} \bra{0}^{\otimes m} U \ket{0}^{\otimes n}\ket{\psi},$$
where $\sqrt{A}$ is a normalization constant and $U$ is a unitary. This is of course only possible if the map $V$ is an (approximate) isometry, which as discussed above requires $n \geq m$.

How do we implement $V$? The obvious first step is to add $n$ ancilla qubits in the state $\ket{0}$, and then apply the unitary $U$. This produces the state
\begin{align}
   U \ket{\psi} \ket{0}^{\otimes n} =  \frac{1}{\sqrt{A}} V \ket{\psi} \ket{0}^{\otimes m} + \sqrt{\frac{A-1}{A}} \sum_{k_{1} \dots k_{m} \neq 0} V_{k_{1} \dots k_{m}} \ket{\psi} \ket{k_1 \dots k_m},
\end{align}
where $V_{k_1 \dots k_m} = \bra{k_1 \dots k_m} U $ selects the $\ket{k_1 \dots k_m}$ component of $U \ket{\psi} \ket{0}^{\otimes n}$. 

\begin{figure}
    \centering
    \includegraphics[width=.45\textwidth]{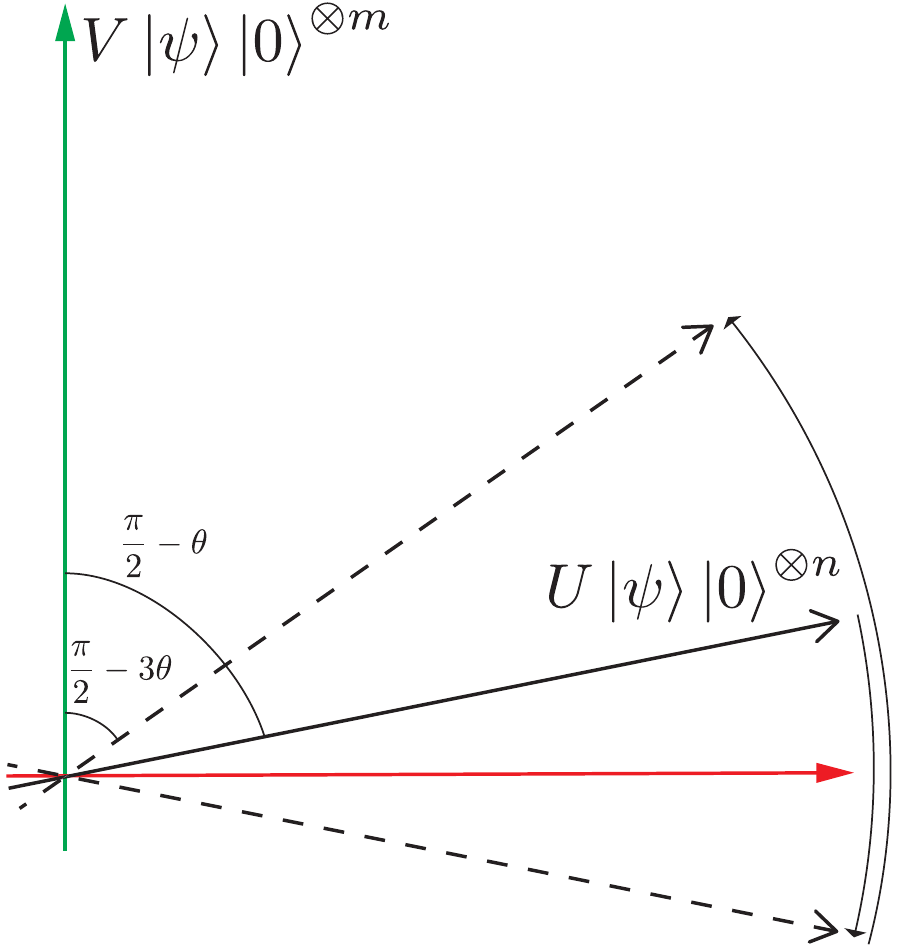}
    \caption{An illustration of the Grover-search algorithm. Acting with $U$ takes $\ket{\psi}\ket{0}^{\otimes n}$ to a state where the $m$ ancilla qubits have a very small probability of being in the desired state $V \ket{\psi}\ket{0}^{\otimes m}$, i.e. at an angle $\pi/2-\theta$ with the green axis with $\theta \sim 1/\sqrt{A}\sim 2^{-m/2}$. One ``back-and-forth'' iteration of the algorithm results in reflections shown by the curved arrows. This sequence of operations moves $U \ket{\psi}\ket{0}^{\otimes n}$ closer to the green axis by an angle $2 \theta$. Therefore to get close to the green axis one needs $\sim 2^{m/2}$ iterations of this procedure, hence an exponential complexity.} 
    \label{fig:grover}
\end{figure}

With the ability to magically postselect onto the $\ket{0}^{\otimes m}$ component, we would now be done. Since we are unable to do that, we instead apply a unitary that adds a phase of $(-1)$ if and only if the $m$ qubits being postselected are all in the state $\ket{0}$. If we consider the two-dimensional subspace spanned by $V \ket{\psi} \ket{0}^{\otimes m}$ and $U \ket{\psi} \ket{0}^{\otimes n}$, for some particular initial state $\ket{\psi}$, as shown in Fig.~\ref{fig:grover}, this acts as a reflection in the horizontal axis~\cite{nielsen2000quantum}.

The next step is to undo the application of $U$, by first applying its adjoint $U^\dagger$, then applying a phase of $(-1)$ if and only if the $n$ ancilla qubits are \emph{not} all still in the state $\ket{0}$, and finally reapplying $U$. On the two-dimensional subspace (see Fig.~\ref{fig:grover}), it can be easily checked that this acts as reflection around the axis generated by $U \ket{\psi} \ket{0}^{\otimes n}$.

We have so far done two reflections in axes separated by an angle $\theta \approx 1/\sqrt{A}$. Combined, these act as a rotation in the two-dimensional space by an angle $2 \theta$. To produce the desired final state $V \ket{\psi} \ket{0}^{\otimes m}$ we need to rotate the state by an angle of $\pi/2$; we do so simply by repeating the above procedure approximately $\pi \sqrt{A}/4$ times.\footnote{This will produce the correct state up to an error of size $O(1/\sqrt{A})$, which is generally very small since $\sqrt{A}$ grows exponentially with the number of qubits being postselected. However there also exist simple tricks for producing the exact output state desired, which is useful when the number of qubits being postselected is $O(1)$. See for example Appendix A of~\cite{BroGha19}.}

For typical scrambling unitaries $U$, we have $\sqrt{A} \sim 2^{-m/2}$, so the total complexity of the entire Grover search algorithm is 
\begin{align}
C = O(\tilde C 2^{m/2}),
\end{align}
where $\tilde C$ is the complexity of implementing the unitary $U$, which needs to be done at each step in the iteration. In tensor networks, $\tilde C$ is generically proportional to the number of tensors in the network. The reconstruction complexity is therefore exponential in the number of postselected qubits $m$.

Of course, Grover search is just one particular quantum algorithm, and it is natural to wonder whether faster algorithms exist. In a black-box setting, Grover search is known to be optimal. It was assumed in~\cite{BroGha19} that this is also true for generic unitaries $U$, even when the algorithm is allowed to depend on $U$.

Even if we accept that Grover search is an optimal reconstruction algorithm for tensor networks, we still need to relate it to actual theories of quantum gravity. The gravitational analogue of a locally minimal cut in a tensor network is a (quantum) extremal surface. The natural analogue of the cut that defines the maximal size of the bulge is also a quantum extremal surface, but it is an extremal surface where, within any Cauchy slice, there exist local perturbations that can decrease the generalized entropy.

A complete gravitational Python's Lunch consists therefore of three quantum extremal surfaces: the minimal QES $\gamma_\text{min}$, a nonminimal (but locally minimal) `appetizer' QES $\gamma_\text{aptz}$ that forms the other end of the lunch, and a third `bulge' QES $\gamma_\text{bulge}$ that sits in between the two and has larger generalized entropy than either. See Fig.~\ref{fig-onelunch}. In other words, we have
\begin{align}
    S_\text{gen}(\gamma_\text{bulge}) > S_\text{gen}(\gamma_\text{aptz}) > S_\text{gen}(\gamma_\text{min}).
\end{align}
Here the generalized entropies $S_\text{gen}  = A/4 G_N + S$ that we are interested in are defined using the maximally mixed state within our code subspace of interest. 

It is worth taking a moment to see why it is the maximally mixed state that plays this privileged role. Recall that, as shown in~\cite{HayPen18, AkeLei19}, whether a given bulk operator admits a (single) boundary reconstruction is determined by the location of the minimal QES for the maximally mixed state. This means that the lunch is only reconstructible if 
\begin{align} \label{eq:reconstructible}
S_\text{gen}(\gamma_\text{aptz}) > S_\text{gen}(\gamma_\text{min})
\end{align}
in the maximally mixed state. We saw exactly the same effect in the tensor network toy model above: for $V$ to be an isometry, we needed the number of legs in $\gamma_\text{aptz}$ to be larger than the number of legs in $\gamma_\text{min}$, plus all the bulk legs in between, so that the output dimension was larger than the input one. This condition is exactly \eqref{eq:reconstructible}, since the generalized entropy of a tensor network cut in the maximally mixed bulk state is equal to the number of legs within the cut plus the number of bulk legs to its right. 

When it comes to the question of reconstruction complexity, the tensor network calculations suggest that what matters is the number of postselected qubits. This is equal to the number of legs in the bulge cut $\gamma_{\text{bulge}}$, plus the number of bulk legs between $\gamma_{\text{bulge}}$ and $\gamma_{\text{aptz}}$, minus the number of legs in the appetizer cut $\gamma_{\text{aptz}}$. Again, this can be naturally rewritten as the difference $S_\text{gen}(\gamma_\text{bulge}) - S_\text{gen}(\gamma_\text{aptz})$ between generalized entropies \textit{in the maximally mixed bulk state}. Assuming that the analogy between tensor networks and gravity continues to hold, it is therefore the generalized entropy of nonminimal extremal surfaces in the maximally mixed state that determines the reconstruction complexity.

Finally, we need to know the gravitational analogue of the size $\tilde C$ of the tensor network. The most natural prescription is that it should be equal to the volume of the maximal volume slice~\cite{StaSus14, Sus14a}, bounded by the minimal QES and the appetizer QES, and potentially restricted to slices containing the bulge QES. An alternative prescription, which tends to give similar answers in practice, is the action of the Wheeler-de Witt patch associated to the lunch~\cite{BroRob15}. Since the complexity depends exponentially on the number of postselected qubits, but only linearly on $\tilde C$, the exact details of the prescription for $\tilde C$ will not be very important for us. In all the examples considered in either~\cite{BroGha19} and this paper, both the volume and action give the same asymptotic scaling for $\tilde C$, and agree with the results from toy models.

The formal statement of the Python's Lunch conjecture for the complexity of decoding bulk operators inside a lunch is as follows that the complexity $C$ of decoding bulk operators in the lunch is given by
\begin{align}
    C = O\left (\tilde C \exp \left[\frac{1}{2}(S_\text{gen}(\gamma_\text{bulge}) - S_\text{gen}(\gamma_\text{aptz})\right]\right).
\end{align}
Note that generalized entropies are defined in base $e$ in accordance with standard convention, and so the base of the exponential is different from when we were working with qubits. It is also worth noting that the complexity $C$ does \emph{not} depend on the generalized entropy $S_\text{gen}(\gamma_\text{min})$ of the minimal QES (although of course we need to have $S_\text{gen}(\gamma_\text{min}) < S_\text{gen}(\gamma_\text{aptz})$ to ensure that it is actually minimal).

\subsection{Multi-Python: breakfast, lunch, and dinner} \label{sec:multiple}
Suppose we have a sequence of bulges: spacelike separated QESs, oscillating in size. Specifically, let the minimal QES be labelled $\gamma_1$, the first bulge surface to its right $\gamma_2$, the first appetizer surface $\gamma_3$, the next bulge $\gamma_4$, etc. 

\paragraph{The Python's Lunch conjecture for multiple bulges:} the restricted complexity $C$ of decoding bulk operators just to the right of $\gamma_1$ is given by
\begin{align} \label{eq:multilunch}
    C = O\left (\max_{j > i} \tilde C_{i,j} \exp \left [\frac{1}{2} (S_\text{gen}(\gamma_j) - S_\text{gen}(\gamma_i)\right]\right).
\end{align}
Here $\tilde C_{i,j}$ is the maximal volume of a partial Cauchy slice bounded by $\gamma_i$ and $\gamma_j$.

To understand this result, it is helpful to first consider two special cases: the first contains two lunches and has $S_\text{gen}(\gamma_5) > S_\text{gen}(\gamma_3)$; the second has $S_\text{gen}(\gamma_5) < S_\text{gen}(\gamma_3)$. Again, we will work explicitly with tensor networks toy models and trust that gravity works analogously.

\begin{figure}
    \centering
    \includegraphics[width=.9\textwidth]{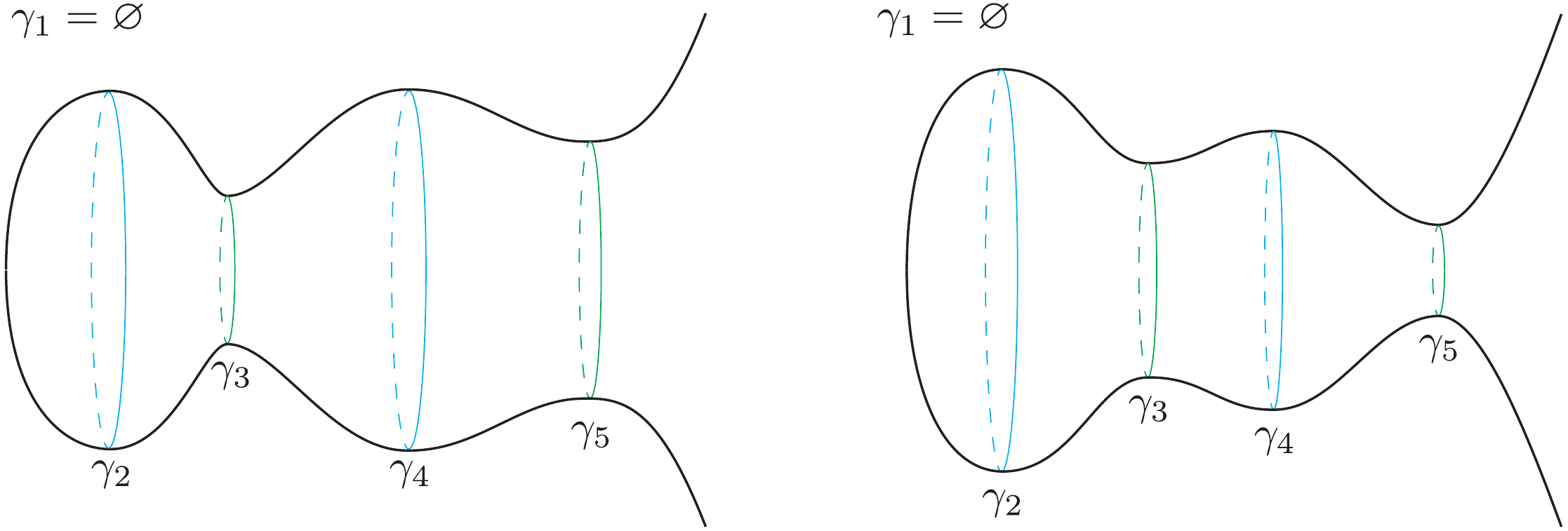}
    \caption{Two types of multiple bulge scenarios are depicted. The salient difference between them is that in the left (right) figure, the throats get bigger (smaller) as we move towards the boundary. For simplicity, the lunches shown are one-sided, i.e. $\gamma_1 = \varnothing$.}
    \label{fig:multilunch}
\end{figure}

In the first case, the maps from $\gamma_1$ to $\gamma_3$ and from  $\gamma_3$ to $\gamma_5$ are both approximate isometries. We can therefore implement the entire isometry by simply implementing one isometry followed by the other. The total circuit complexity is the sum of the complexities associated to each step, which will generally be dominated by the larger of the two. We therefore find that the complexity scales as $$O\left (\exp\left[ \frac{1}{2} \max(S_\text{gen}(\gamma_\text{2}) - S_\text{gen}(\gamma_\text{3}),S_\text{gen}(\gamma_\text{4}) - S_\text{gen}(\gamma_\text{5}))\right]\right).$$ This is consistent with \eqref{eq:multilunch}.

In the second case, the map from $\gamma_3$ to $\gamma_5$ is no longer an isometry, because there are more qubits postselected than ancilla qubits added. If $n_i \propto  S_\text{gen}(\gamma_i)$ is the effective number of qubits associated to each cut $\gamma_i$, then there are $n_4 - n_5$ qubits postselected, but only $n_4 - n_3$ ancilla qubits added. It is helpful to break this map down into two pieces: an isometry where $n_4 - n_3$ ancilla qubits are added, a unitary is applied and then $[n_4 - n_3 - O(1)]$ qubits are postselected, and an additional postselection of $[n_3 - n_5 + O(1)]$ qubits. The first step can be implemented unitarily without any knowledge of what the tensor network looks like to the left of $\gamma_3$, with circuit complexity $O(2^{(n_4 - n_3)/2})$.

To go further, we need to take advantage of the fact that the input at $\gamma_3$ is the output of an isometry from $\gamma_1$ with circuit complexity $O(2^{(n_2 - n_3)/2})$. So the entire isometry $V$ from $\gamma_1$ to $\gamma_5$ can be written as
\begin{align}
    V \ket{\psi} \propto \bra{0}^{n_3 - n_5 + O(1)} U \ket{0}^{n_3 - n_1 + O(1)} \ket{\psi},
\end{align}
for a unitary matrix $U$ with circuit complexity $O(2^{(n_4 - n_3)/2}) + O(2^{(n_2 - n_3)/2})$. We can therefore implement $V$ using Grover search at the cost of implementing the unitary $U$ (and its inverse) $O(2^{(n_3 - n_5)/2})$ times. This gives a final answer for the complexity that scales as $$O(\exp[\frac{1}{2} \max(S_\text{gen}(\gamma_\text{2}) - S_\text{gen}(\gamma_\text{5}),S_\text{gen}(\gamma_\text{4}) - S_\text{gen}(\gamma_\text{5}))]).$$ Again, this is consistent with \eqref{eq:multilunch}.

Having understood these two examples, the general rule is fairly simple to derive. As long as each locally minimal cut is larger than the one before, each isometry can be implemented in turn and the dominant contribution comes from the step with largest complexity. However, when a locally minimal cut is smaller than previous ones, the only way to implement the additional postselection is to do a Grover search that goes all the way back to the first smaller locally minimal cut. This additional complexity therefore multiplies the complexities of all the intermediate steps that are used in that Grover search. The final answer for the total complexity is therefore exponential in the largest \emph{total net decrease} in generalized entropy between any two surfaces when moving through the network, as claimed in \eqref{eq:multilunch}.

\section{An Unexpected Python} \label{sec:maincalc}

\begin{figure}
    \centering
    \includegraphics[width=0.55\textwidth]{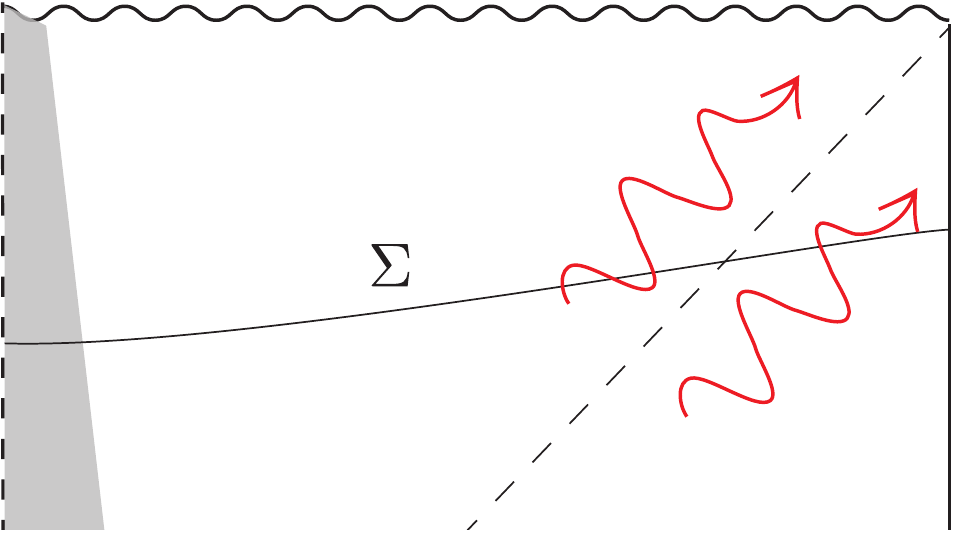}
    \caption{The late time Cauchy slice $\Sigma$ in an isolated black hole formed from collapse. We construct our code subspace from a Rindler decomposition of the Hawking modes on $\Sigma$.}
    \label{fig:late_time}
\end{figure}

In this section, we will describe the code subspace $\mathcal{H}_{\text{code}}$ associated to the state of the horizon in a large single-sided non-evaporating AdS black hole (e.g. formed by collapse) at late times and show that there is a QES in its maximally mixed state. As noted in the introduction, we consider a subspace of states allowing arbitrary excitations of outgoing Rindler-like modes (regulated away from the event horizon) straddling the event horizon in the late time adiabatic regime. In particular, we work on a late time Cauchy slice $\Sigma$ and focus on a code subspace defined by a Rindler decomposition near the event horizon on $\Sigma$; see Fig.~\ref{fig:late_time}. The state that we will be interested in is one of reduced entanglement between Hawking partners compared with the Hartle-Hawking (HH) state: the interior and exterior Hawking modes on $\Sigma$ will together be in a mixed state. This choice of state in the code subspace defined on $\Sigma$ affects the spacetime to the past of $\Sigma$, generating a large blueshift. It is precisely this effect (both on the state of the quantum fields and its backreaction on the geometry) that gives rise to the novel, nonminimal QES  that is not present for the equilibrium state. Note that the maximally mixed state is able to evade the argument against the existence of a nontrivial QES from Sec. \ref{sec:intro} because, unlike its equilibrium counterpart, it has a white hole singularity; thus the QES is not causally separated from the asymptotic boundary.

Since our purpose here is to illustrate the phenomenon rather than provide a general proof thereof, we will permit ourselves several simplifying assumptions: we will model the bulk theory via a free scalar field or a free fermionic field on a black hole background that at late times approaches a large Schwarzschild-AdS. We expect that the story is morally unchanged upon adding spin to the black hole or considering other bulk quantum fields, and we will comment on this in Sec.~\ref{sec:disc}. 

We will therefore take the late time behavior of our spacetime to be well-described by a portion of Schwarzschild-AdS$_{d+1}$ in Kruskal coordinates:
\begin{align}
    ds^2 = -\frac{1}{4 \pi^2 T^2} \frac{F(r)}{UV} dU dV + r^2 d \Omega^2,
\end{align}
where $T$ is the black hole temperature and $F(r)$ is the emblackening factor:
\begin{align}
    F(r) = 1-\frac{\mu}{r^{d-2}}+\frac{r^2}{\ell_{\text{AdS}}^2},
\end{align}
here $\ell_{\text{AdS}}$ denotes the AdS radius and $\mu$ is a constant related to the black hole mass. In the near horizon regime where the metric is given by
\begin{align}
    ds^2 = -2 dU dV + r(U,V)^2 d \Omega^2 + O(UV),
\end{align}
with
\begin{align}
    r(U,V) = r_s + 2\pi T~U V + O(UV),
\end{align}
with $r_s$ the horizon radius and $T$ the black hole temperature. In this regime, the state of the bulk quantum fields approximately factorizes into infalling and outgoing modes. The infalling sector admits a simple reconstruction from boundary data via time-reversed evolution. The outgoing sector is more mysterious: backwards evolution is of little use due to the blueshift in the outgoing modes and resultant transplanckian problem. As alluded to above, this blueshift has a critical role to play: below we demonstrate that in the maximally mixed state of $\mathcal{H}_{\text{code}}$ those very same blueshifts result in a new outermost quantum extremal surface $\gamma_{\text{aptz}}$. The existence of this QES  -- which is nonminimal -- causes the reconstruction of the interior outgoing modes to be exponentially complex, as discussed in Sec.~\ref{sec:PLreview}.


\subsection{Restricted maximin} \label{sec:resmaximin}

Rather than directly solving for the location of the nonminimal QES $\gamma_{\mathrm{aptz}}$, we will indirectly prove its existence and approximate its location by finding surfaces where the quantum expansion~\cite{Wal10QST, BouFis15a} has a particular sign.

We begin by defining the quantum expansion. Let $\sigma$ be a smooth codimension-2 surface homologous to the entire boundary with homology hypersurface $H_\sigma$. The quantum expansion~\cite{Wal10QST, BouFis15a} of the null congruence generated by the outwards-future-directed null vector field $k^{a}$ normal to $\sigma$ is defined in the following way: pick $\lambda$ to be an affine parameter along $k^{a}$ such that $\sigma$ is at $\lambda = 0$. Now, let $\sigma_V$ be the surface $\lambda = V(y)$ where $y$ denotes the transverse direction. Then
\begin{align}\label{eq-Thetakdef}
    \Theta_{k}[\sigma; y] = \frac{4 G_N}{\sqrt{h(y)}} \left.\frac{\delta S_{\text{gen}} [H_{\sigma_V}] }{\delta V(y)}\right\rvert_{V=0},
\end{align}
where $h(y)$ denotes the determinant of the intrinsic metric on $\sigma$. Similarly, for a null congruence generated by the inwards-future-directed $\ell^{a}$ fired from $\sigma$ (with $V(y)$ replaced by $U(y)$ for clarity), we have
\begin{align}\label{eq-Thetaldef}
        \Theta_{\ell}[\sigma; y] = \frac{4 G_N}{\sqrt{h(y)}} \left.\frac{\delta S_{\text{gen}} [H_{\sigma_V}] }{\delta U(y)}\right\rvert_{U=0}.
\end{align}

The condition that indirectly implies the existence of $\gamma_{\mathrm{aptz}}$ is the presence of a surface $\sigma$ satisfying
\begin{align} \label{eq-Thetak}
    &\Theta_{k}[\sigma]\leq 0,\\
    &\Theta_{\ell}[\sigma] \geq 0, \label{eq-Thetal}
\end{align}
where by dropping the $y$ label we mean that the condition applies at all points on $\sigma$. Since $\sigma$ is the opposite of a ``normal'' surface, in which the quantum expansion expands towards the exterior and contracts towards the interior, we call it ``quantum anti-normal''. Using the quantum focusing conjecture~\cite{BouFis15a}, it was shown in~\cite{BroGha19, BouSha21}  that the existence of a quantum anti-normal surface is sufficient to guarantee the existence of a QES in $D[H_{\sigma}]$. The crux of this argument is the restricted quantum maximin prescription~\cite{Wal12, MarWal19, AkeEng19b}, in which a QES is found by minimizing $S_{\mathrm{gen}}$ over all surfaces on a given Cauchy slice, and then maximizing the minimal $S_{\text{gen}}$ over all Cauchy slices. The intuition is roughly that the quantum maximin surface in $D[\Sigma]$ cannot intersect $\sigma$ because~\eqref{eq-Thetak} and~\eqref{eq-Thetal} ensure that moving away from $\sigma$ lowers the generalized entropy; a similar argument then shows that $\sigma$ cannot intersect the boundary of $D[H_{\sigma}]$. The maximin surface must then lie in the interior of $D[H_{\sigma}]$ and hence is quantum extremal.

Finding a surface satisfying~\eqref{eq-Thetak} in the interior of a stationary black hole is easy; The expansion $\Theta_{k}$ vanishes on the horizon and decreases as we move inwards along spheres. When the future horizon is in the HH state, $\Theta_{\ell}\leq 0$ for spheres near the horizon; this follows from the fact that the classical expansion $\theta_{\ell}$~\footnote{The classical expansion can be obtained by substituting $4G_{N} S_{\text{gen}}$ by $A$ in the definition of the quantum expansion in \eqref{eq-Thetakdef} and \eqref{eq-Thetaldef}.} is negative and $O(G_{N}^0)$ while the quantum correction is subleading. However, as we establish below, our choice of mixed state in $\mathcal{H}_{\text{code}}$ has large von Neumann entropy variations in the $\ell^{a}$ direction. Moreover, these get enhanced due to the near-horizon blueshift of the horizon modes as they evolve backwards, which yields precisely the inequality that we need.

\begin{figure}
    \centering
    \includegraphics[width=0.5\textwidth]{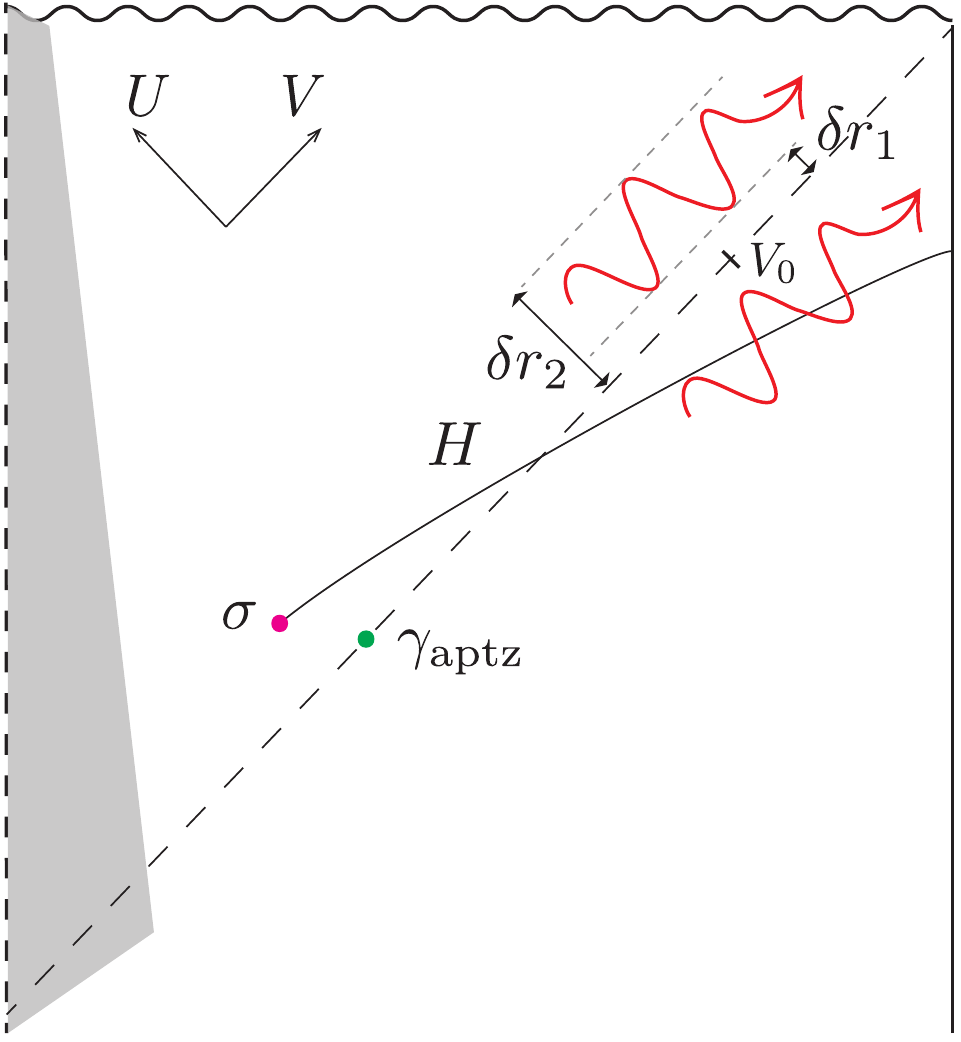}
    \caption{A spherically symmetric black hole formed by collapse. Our code subspace $\mathcal{H}_\text{code}$ allows arbitrary excitations of outgoing Hawking wave-packet partners localized to some range of $U$. At some point $V_0$ along the horizon, this range can be specified by small $\delta r_1$ and $\delta r_2$ satisfying $\delta r_2 /\delta r_1 \gg 1$. The part of this region in the black hole interior (referred to by $\Delta$ in the main text) is depicted by fine dashed lines. If $V_0$ is of order a scrambling time to the future of the last infalling matter, in the maximally mixed or thermal state of $\mathcal{H}_\text{code}$ there exists a quantum anti-normal surface $\sigma$ but about a scrambling time in the past of $V_0$. The existence of $\sigma$ then indirectly implies the existence of a quantum extremal surface $\gamma_{\text{aptz}}$ in the exterior of $\sigma$; to leading order in $G_{N}$, $\gamma_{\text{aptz}}$ is on the horizon.}
    \label{fig:disentangle}
\end{figure}

\subsection{The code subspace}
As noted above, we construct the code subspace explicitly for massless free scalars and also for fermion fields. For simplicity we will focus here on the spherically symmetric modes close to the horizon. We will discuss the non-spherically symmetric case in Sec. \ref{sec:disc}. In the HH state, the outgoing Rindler-like modes are in the following state:
\begin{align}\label{eq-Omega}
    \bigotimes_{\omega} N_\omega \sum_{n=0}^{\infty \text{ or }1}e^{-\pi \omega n} \ket{n}^{\omega }_{\text{in}}\ket{n}^{\omega}_{\text{out}},
\end{align}
where $\ket{n}^{\omega }_{\text{in}}$ and $\ket{n}^{\omega }_{\text{out}}$ denote the state of definite Rindler frequency $\omega$ in the interior and exterior respectively. Note that the sum goes up to infinity for a scalar and to $1$ for a fermion. $N_\omega$ is a normalization factor equal to $(1-\exp(-2\pi \omega))^{1/2}$ or $(1+\exp(-2\pi \omega))^{-1/2}$ for the scalar and fermion respectively.

We can \emph{almost} specify our code subspace $\mathcal{H}_{\text{code}}$ as the span of states of the form $\ket{n}^{\omega }_{\text{in}}\otimes \ket{m}^{\omega }_{\text{out}}$. However, states of definite Rindler frequency have divergent energies at the horizon. A more physical $\mathcal{H}_{\text{code}}$ can be constructed from wave-packets that are localized a small neighborhood away from the horizon.

Let us describe these localized wave-packets in more detail. At some $V=V_0$ on the horizon, we define a range $\Delta$ of $U$ corresponding to spheres between radii $r_s-\delta r_1$ and $r_s-\delta r_2$ such that $0<\delta r_1 \ll \delta r_2 \ll r_s$. More directly, $\Delta$ is given by $\epsilon<U<\epsilon e^L$ for

\begin{align}\label{eq-epsilonL1}
    \epsilon =& \frac{\delta r_1}{2\pi T V_0},\\
    \epsilon e^L =& \frac{\delta r_2}{2\pi T V_0}\label{eq-epsilonL2},
\end{align}
where $\epsilon V_0 \ll 1$ and $L \gg 1$ such that $\epsilon e^L V_0 \ll 1$. See Fig.~\ref{fig:disentangle} for an illustration.

Modes restricted to $\Delta$ will not have the exact entanglement structure of \eqref{eq-Omega}. However, we can construct wave-packets with mean Rindler frequency $\omega$ which are localized to $\Delta$.\footnote{Such wave-packets have a Rindler frequency variance of order $1/L$.} By picking evenly spaced $\omega$ within some range $[\omega_0 - \delta \omega/2, \omega_0 + \delta \omega/2]$ with $\delta \omega \ll\omega_0$, we can find approximately $h \approx \delta \omega L/ 2 \pi$ orthogonal wave-packets within $\Delta$. Let $\widetilde{\ket{n}^{\omega}_{\text{ in}}}$ and $\widetilde{\ket{n}^{\omega}_{ \text{out}}}$ denote the state with occupation number of this interior wave-packets and their outside partners. In the HH state, we have:
\begin{align}
\ket{\Omega} \approx \bigotimes_{j=1}^{h} N_\omega \sum_{n=0}^{\infty \text{ or }1} e^{-\pi \omega_j n} \widetilde{\ket{n}^{\omega_j}_{\text{in}}}\widetilde{\ket{n}^{\omega_j}_{\text{out}}},
\end{align}
where
\begin{align}
  \omega_j = \omega_0 -\frac{\delta \omega}{2} + j \frac{2\pi}{L}.  
\end{align}
Our $\mathcal{H}_{\text{code}}$ to which $\ket{\Omega}$ belongs is:
\begin{align}
    \mathcal{H}_{\text{code}} = \bigotimes_{j=1}^{h} \widetilde{ \mathcal{H}^{\omega_j}_{\text{in}}} \otimes \widetilde{\mathcal{H}^{\omega_j}_{\text{out}}},
\end{align}
where $\widetilde{\mathcal{H}^{\omega}_{\text{in}}}$ and $\widetilde{\mathcal{H}^{\omega}_{\text{out}}}$ denote the Hilbert spaces of the interior and exterior wave-packets, respectively.

Since our code subspace is formally infinite dimensional for bosonic modes, it does not have a true maximally mixed state. We could avoid this problem by restricting our code subspace to states with an occupation number $n$ below some arbitrary upper bound. Alternatively we could restrict ourselves to fermionic modes, which are inherently finite dimensional. 

However, the easiest and most natural approach is to regulate the infinity by simply working with a thermal state $\rho$, which on this code subspace is:
\begin{align}\label{eq-mixedstate}
\rho = \bigotimes_{j=1}^{h} N_{\omega_j}^2\left( \sum_{n=0}^{\infty \text{ or }1} e^{-2\pi \omega_j n} \widetilde{\ket{n}^{\omega_j}_{\text{in}}} \widetilde{\bra{n}^{\omega_j}_{\text{in}}} \right) \left( \sum_{n=0}^{\infty \text{ or }1} e^{-2\pi \omega_j n} \widetilde{\ket{n}^{\omega_j}_{\text{out}}} \widetilde{\bra{n}^{\omega_j}_{\text{out}}} \right),
\end{align}
where the in and out wave-packets have both been sufficiently disentangled that one contains essentially no usable information about its counterpart (note that the differences between $\rho$ and $\ket{\Omega}\bra{\Omega}$ come from the $s$-wave sector). Essentially, using a thermal state rather than a maximally mixed state corresponds to using a Bayesian prior when reconstructing the modes where high occupation numbers for the Rindler-like modes are assumed to have exponentially suppressed probability.

It may seem worrying that the state $\rho$ backreacts strongly on the geometry in the far past due to the near-horizon blueshift, in apparent tension with formulating a code subspace. The resolution is in the proper choice of bulk slice. The intrinsic geometry of a ``boosted’’ slice (similar to $H$ in Fig.~\ref{fig:disentangle}) will be the same to leading order in $G_N$ among various states of $\mathcal{H}_{\text{code}}$. We will in addition make the assumption that the region behind the horizon has exactly the same geometry in $\rho$ as in the HH state. This assumption holds with a particular choice of gravitational dressing where the state $\rho$ is prepared from the HH state by the action of random unitaries on the outside Hawking partners. This will decohere the partner modes and create the state $\rho$, but by bulk locality manifestly preserves the geometry behind the horizon.\footnote{We expect that other gravitational dressings also lead to a new quantum extremal surface, but demonstrating it rigorously requires careful analysis of the backreaction.}

We now proceed to computing the quantum expansion of spheres near the horizon in the state $\rho$.
\subsection{The quantum expansion}
Since we are restricting to spherically symmetric spacetimes, we only consider the generalized entropy of the exterior of a sphere at the location $(U,V)$:
\begin{align}
    S_{\text{gen}} [H (U,V)] = \frac{A(U,V)}{4 G} + S [H (U,V)],
\end{align}
where $A$ denotes the area of the $(U,V)$ sphere and $S$ denotes the von Neumann entropy of a homology hypersurface $H (U,V)$ outside of the sphere (note that, by contrast with a BH coupled to a reservoir, this is independent of our choice of $H(U,V)$ because the boundary conditions are reflecting). The corresponding quantum expansions are:
\begin{align}\label{eq:kruskalmetric}
    \Theta_{V}(U,V) &= \frac{4G_{N}}{A(U,V)}\frac{\partial S_{\text{gen}}[H(U,V)]} {\partial V},\\
    \Theta_{U}(U,V) &= \frac{4G_{N}}{A(U,V)}\frac{\partial S_{\text{gen}}[H(U,V)]} {\partial U}. \label{eq-ThetaU}
\end{align}
In the global (AdS) HH state, the spacetime isometries guarantee stationarity of the bifurcate horizon:
\begin{align}
\Theta_{U}(U, V=0)=0, \label{eq:V=0thetaU} \\
\Theta_{V}(U=0, V) =0. \label{eq:U=0thetaV}
\end{align}
In large AdS black holes formed by collapse and then allowed to equilibrate, \eqref{eq:U=0thetaV} holds as stated, while \eqref{eq:V=0thetaU} survives as the statement that \begin{align}
\Theta_{U}(U,V) = O(V)
\end{align}
in the limit of small $V$ and fixed $U$. This is because the classical spacetime and the reduced state of the bulk quantum fields are the same for the AdS HH state restricted to $V > 0$ and the post-collapse black hole.

Recall that our goal here is not to find a QES explicitly but rather to find a surface $\sigma$ with negative $\Theta_{V}$ and positive $\Theta_{U}$, which as noted above is sufficient to guarantee the existence of a QES between $\sigma$ and the asymptotic boundary $\mathscr{I}$. We will discuss the variations of the area and entropy separately. In the near horizon region the area of spheres is
\begin{align}
    A(U,V) \approx \text{Vol}(\mathbb{S}^{d-1}) r_s^{d-1} (1 -2~U V) + o(UV),
\end{align}
where $\text{Vol}(\mathbb{S}^{d-1})=2\pi^{d/2}/\Gamma(d/2)$ is the area of a $(d-1)$ dimensional unit sphere. Therefore the classical expansion $\theta_V = A(U,V)^{-1} \partial_{V} A(U,V)= -2 U + o(U)$. It is negative in the interior; $\theta_U = -2V + o(V)$ is likewise negative.

The bulk entropy term $S [\Sigma (U,V)]$ is difficult to compute exactly, but under reasonable simplifying assumptions we may approximate it to an accuracy sufficient for establishing the existence of a quantum anti-normal sphere. The state $\rho$ tensor factorizes across different angular momenta by construction, which by spherical symmetry have decoupled dynamics. The contribution to the von Neumann entropy from high angular momentum modes includes a divergent term proportional to $A(U,V)$, which is balanced by a renormalization of $G_{N}$ in $S_{\text{gen}}$~\cite{SusUgl94, Kab95, LarWil95, Jac94, FurSol94}. As noted above, the change in the state from $\Omega$ to $\rho$ involves a change in the s-wave sector only, so we may focus exclusively on the entropy contribution of the s-wave sector.

To find a quantum anti-normal surface, we take a fixed $U > 0$ (within the range $\Delta$) where the Rindler modes were disentangled, and then take the limit of small $V$.

We first want to to show $\Theta_V <0$. Since the classical expansion $\theta_V$ is already negative, this is not especially surprising, but we still need to be somewhat careful in case the bulk entropy gradient $\partial_V S$ is singular in the $V \to 0$ limit. Indeed, if the infalling modes were at a different temperature $T_\text{in}$ from the black hole temperature $T$ then this is exactly what would happen: the (renormalized) bulk entropy gradient is \cite{Pen19}
\begin{align}
    \partial_V S \sim \frac{1}{V}\left[1 - \frac{T_\text{in}}{T}\right] + O(V^0),
\end{align}
which is clearly singular in the limit of interest. However, because our black hole is in equilibrium, this issue does not arise and $\partial_V S$ is well behaved. Essentially this is because we would have the same bulk state (when restricted to the region of interest) if we had started with the two-sided HH state (rather than a black hole formed from collapse) and then disentangled the outgoing wave-packets as usual. The resulting state is manifestly nonsingular at $V=0$, as is its entropy gradient. It follows that we have
\begin{align}
    \Theta_{V}(U, V \to 0) = -2U + \frac{4 G_{N}}{A} \frac{\partial S}{\partial V}  = -2U +  O(G_{N}) < 0
\end{align} 
for any fixed $U > 0$ in the limit $G_{N} \to 0$.

We emphasize that $\partial_V S$ can still be significantly affected (even at very small $V$) by disentangling the outgoing modes, despite the limit necessarily remaining nonsingular, because the ingoing and outgoing modes are related by the reflecting boundary conditions. An analogous phenomenon was found in \cite{AlmMah19b} where an equilibrating two-sided black hole coupled to a bath had a nonsingular but nonzero entropy gradient at the bifurcation surface. Note that by picking $U$ to be in the range $\Delta$ we ensure that $\theta_V$ is $O(G_{N}^0)$ and negative, upholding $\Theta_V <0$ despite any non-zero $\partial_V S \sim O(1)$. 

It only remains to show the somewhat more surprising fact that we also have $\Theta_{U} > 0$ in the same region. This can be derived in two complementary ways: first, by a direct computation of the entropy of quantum harmonic oscillators, and second, by a conformal mapping to Minkowski space. We describe each method in turn. 

\paragraph{Direct Computation:} In the HH state, the symmetries guarantee that $\partial_U S$ goes to zero as $V$ goes to zero. On grounds of dimensional analysis, it is natural to expect that $\partial_U S = O(V)$. In appendix \ref{sec:JTcalc}, we study the entropy variations in JT gravity and explicitly derive this dependence. The intuition behind this term is the mixing between the outgoing and infalling modes in the dynamics away from the horizon, e.g. reflecting asymptotic boundary conditions; this causes some purifications of the near horizon interior out-movers at $V$ to miss $H(U=0,V_0)$. In the state $\rho$, we expect a large correction to $\partial_U S$: this is precisely the entropy derivative that notices the change in the entanglement structure of the out-movers. We can estimate this by comparing the entanglement structure of $\rho$ with that of the HH state, i.e. by considering $\partial_U (S^\rho - S^{HH})$.

Let us be precise about this. Each Rindler wave-packet with mean frequency $\omega$ is a quantum harmonic oscillator at temperature $1/2\pi$ with entropy $S_{\text{th}}(\omega)$ where for the free scalar
\begin{align}\label{eq-entropymodeB}
S_{\text{th}}(\omega) = \frac{2 \pi \omega}{e^{2\pi\omega}-1}- \log ( 1 - e^{-2\pi \omega}),
\end{align}
and for the free fermion
\begin{align}\label{eq-entropymodeF}
S_{\text{th}}(\omega) = \frac{2 \pi \omega}{e^{2\pi\omega}+1}+ \log ( 1 + e^{-2\pi \omega}).
\end{align}

Hence, compared with the HH state, there is a deficit of approximately $2 h S_{\text{th}}(\omega_0)$ in the amount of entanglement with the exterior region.

Since the entanglement pattern of s-wave Hawking modes across the horizon is local on a logarithmic $U$ scale, the logarithmic derivative  $U\partial_U S$ is approximately constant across $\Delta$. Together with the knowledge of the total entanglement deficit this implies that for $U$ in $\Delta$:
\begin{align}\label{eq-dUS}
    U \frac{\partial S[\Sigma (U,V_0)]}{\partial U} = \frac{\delta \omega}{\pi} S_{\text{th}}(\omega_0) + o\left((UV)^0\right).
\end{align}

\paragraph{Mapping to Two Minkowski Spaces:} An alternative way to derive \eqref{eq-dUS} is via a conformal transformation inside and outside of the Rindler regions into two copies of Minkowski space with the corresponding Minkowski null coordinates:
\begin{align}
u_{\text{in}} &= \log U\\
u_{\text{out}} &= -\log (-U).
\end{align}
The requisite conformal factor is $e^{-u}$, and the Minkowski regions are in the thermofield double state with temperature $1/2\pi$. Let $S_{\text{Mink}}(u)$ denote the entropy of the union of the outside Minkowski system with the $u_{\text{in}}<u$ region of the inside Minkowski system. On scales much larger than $2\pi$, the local entanglement structure of the thermal state gives the constant $\partial_u S_{\text{Mink}}(u) = -c/6$ (where $c=1$ for a free scalar and $c=1/2$ for a free fermion). To convert this back to the Rindler answer, we need to also add the contribution coming from the rescaling of the UV cutoff in the universal divergence $c/6 \log (1/\epsilon(u))$, where $\epsilon(u)=\epsilon_0 e^{-u}$. The contribution from this term, $c/6 \partial_u \log (1/\epsilon(u)) = c/6$, precisely cancels the previous contribution, giving $\partial_u S = 0$.

\begin{figure}
    \centering
    \includegraphics[width=0.8\textwidth]{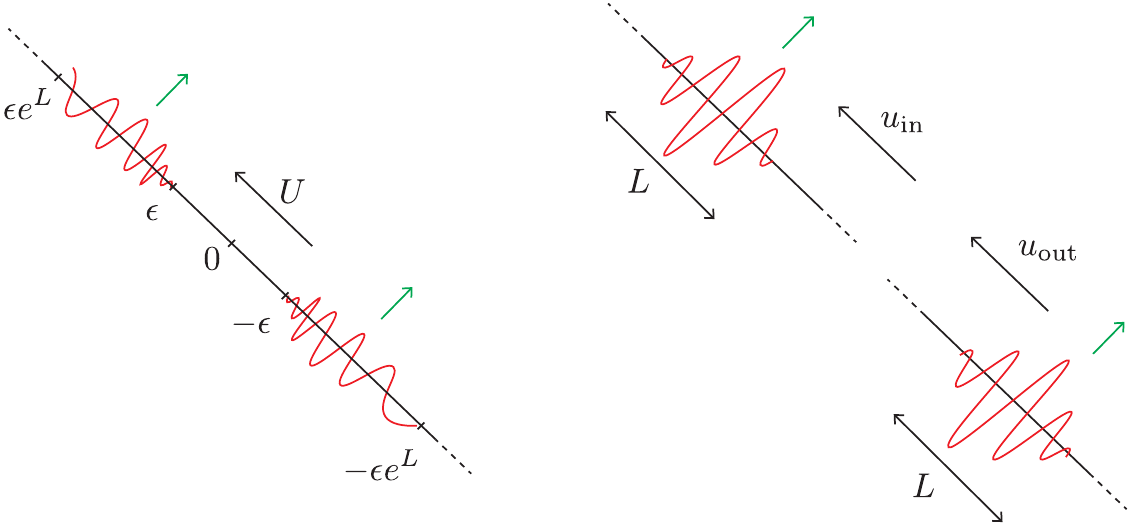}
    \caption{Two scenarios related by a Weyl transformation are shown. On the left we have two right-moving Rindler wave-packet partners within some range of Minkowski null coordinate $U$, i.e. $\epsilon <|U|<\epsilon e^{L}$. On the right we have two separate Minkowski spaces in the thermofield double state of temperature $1/2\pi$. The dual wave-packets will be right-moving Minkowski wave-packets of width $L$. }
    \label{fig:RindMink}
\end{figure}

In the Minkowski picture, the disentangled state $\rho$ is almost in the same thermofield double state except that Minkowski wave-packets localized to $\log \epsilon<u_{\text{in}}< L+ \log \epsilon$ and $\log \epsilon<u_{\text{out}}< L+ \log \epsilon$ are in separate mixed states instead of being entangled with each other. This changes the constant entropy gradient to $\partial_u S_\text{Mink} = -1/6 + (\delta \omega / \pi) S_{\text{th}}(\omega_0)$. Taking the cutoff rescaling into account, we find
\begin{align}
\partial_u S = \frac{\delta \omega}{\pi} S_{\text{th}}(\omega_0).
\end{align}
Changing back to Kruskal coordinates gives \eqref{eq-dUS}.

\vspace{0.4cm}

\noindent For some $U$ in the range $\Delta$ and $V\leq V_0$, we therefore obtain
\begin{align}
\Theta_U (U,V) = -2 V +\frac{4 G_{N} \delta\omega S_{\text{th}}(\omega_0)}{\pi\text{Vol}(\mathbb{S}^{d-1})~U r_s^{d-1}} + o(V, {G_N}^0 V^0).\label{eq-ThetaUfinal}
\end{align}
Here the first term comes from the area term, while the second term comes from the bulk entropy gradient \eqref{eq-dUS}. Since the first term on the right hand side decreases as we decrease $V$, we ought to find $\Theta_U >0$ for some $V_1 < V_0$. To find the parametrically largest such $V_1$, we can take the limit where $U$ is closest to the horizon while still in $\Delta$, i.e. $U \to \delta r_1/ 2\pi T V_0$. Then, the following condition guarantees $\Theta_U >0$:
\begin{align}\label{eq-V1}
\frac{V_1}{V_0} \lesssim \frac{G_N T \delta\omega S_{\text{th}}(\omega_0)}{r_s^{d-1} \delta r_1}.
\end{align}
Therefore at $(U_0, V_1)$, we have a sphere $\sigma$ (see Fig.~\ref{fig:disentangle}) satisfying conditions \eqref{eq-Thetak} and \eqref{eq-Thetal}. Note that $V_1$ is approximately a scrambling time to the past of $V_0$.

As previously discussed in Sec.~\ref{sec:resmaximin}, the restricted quantum maximin prescription implies the existence of a quantum extremal surface $\gamma_{\text{aptz}}$ in the outer wedge of $\mu$. The surface $\gamma_{\text{aptz}}$ is a sphere at $V$ with $V/V_0$ of the same order as $V_1/V_0$. For larger $V$, the area derivative term is parametrically larger than the entropy derivative. Additionally, to satisfy $\Theta_V = 0$, we need $U = O(G)$, or else the area derivative is parametrically large and cannot be balanced by the entropy derivative.

\begin{figure}
    \centering
    \includegraphics[width=0.55\textwidth]{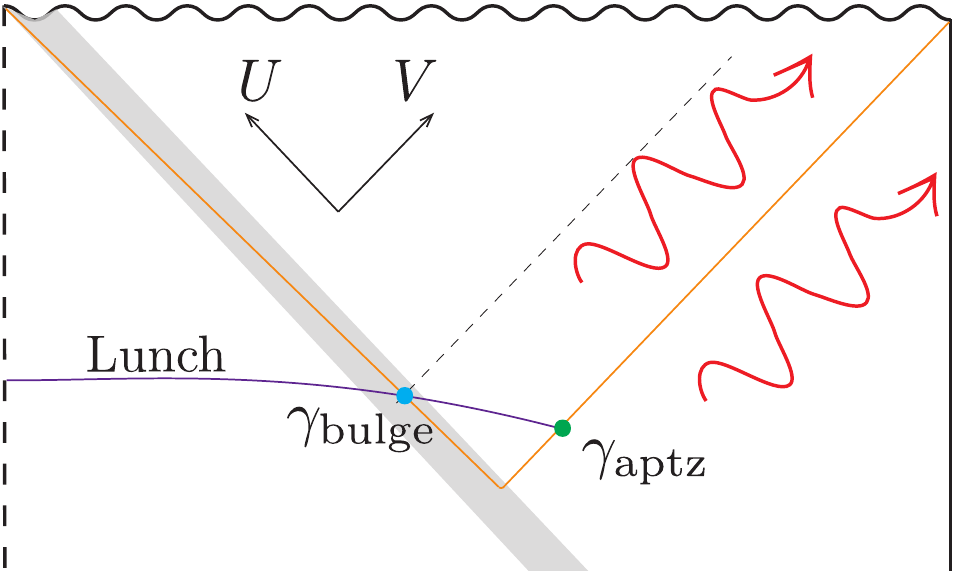}
    \caption{The Penrose diagram of a lunch in a non-evaporating spherically symmetric AdS black hole formed by the collapse of a thin null shell (shown in grey). The solid orange line is foliated by spherical apparent horizons. The mixed state of the code subspace of the Hawking wave-packets results in $\gamma_{\text{aptz}}$ and $\gamma_{\text{bulge}}$ quantum extremal surfaces. The lunch region is the interior of $\gamma_{\text{aptz}}$, a slice of which is shown in purple. Since $\Theta_V=0$ requires small $\theta_V$, both extremal surfaces will be on an apparent horizon to leading order in $G_{N}$.}
    \label{fig:lunch2}
\end{figure}


\subsection{The size of the lunch}
To characterize the lunch, we also need to find the maximal QES $\gamma_{\text{bulge}}$, whose existence follows from the presence of $\gamma_{\text{aptz}}$ by the maximinimax prescription~\cite{BroGha19}. Maximinimax locates $\gamma_{\text{bulge}}$ in the inner wedge of $\gamma_{\text{aptz}}$ but does not narrow down its location further. To get an idea of the location, let us specialize to the AdS-Vaidya collapsing null shell. Since entropy derivatives in the state $\rho$ are small in the $V$ direction, $\Theta_V=0$ can only happen on spheres that satisfy $\theta_{V}=o(G_{N}^0)$. On the other hand, since all spheres satisfy $\theta_U=O(G_{N}^0)$, $\Theta_U=0$ requires highly blueshifted bulk entropy density. The only location which satisfies all of these criteria is inside the Vaidya shock where the apparent horizon intersects the past of the interior wave-packet (See Fig.~\ref{fig:lunch2}).

We can now estimate $S_{\text{gen}}(\gamma_{\text{bulge}}) - S_{\text{gen}}(\gamma_{\text{aptz}})$, the determining factor in the predicted complexity of reconstructing the code subspace. Here it is useful to compare the mixed state $\rho$ with the vacuum state at the horizon. In both states, the region outside of $\gamma_{\text{aptz}}$ is in the same state and therefore has the same generalized entropy.\footnote{Of course even though $\gamma_{\text{aptz}}$ exists in both states it is only quantum extremal in the state $\rho$.} However, compared to the vacuum state, the region outside of $\gamma_\text{bulge}$ has gained the entropy associated with the disentangled wave-packets in $\mathcal{H}_{\text{code}}$. This implies that~\footnote{Note that the fluctuations in the size of the lunch are suppressed compared to the size by $1/\sqrt{h}$.}
\begin{align}\label{eq-difference}
    S_{\text{gen}}(\gamma_{\text{bulge}}) - S_{\text{gen}}(\gamma_{\text{aptz}}) \approx 2 h S_{\text{th}}(\omega_0).
\end{align}
By the Python's Lunch conjecture, this predicts a reconstruction complexity exponential in the size of $\mathcal{H}_{\text{code}}$. An alternative way to derive \eqref{eq-difference} is by integrating \eqref{eq-ThetaUfinal} across the range $\Delta$.

In this analysis we have ignored the area difference between $\gamma_{\text{aptz}}$ and $\gamma_{\text{bulge}}$ because its contribution to $S_{\text{gen}}(\gamma_{\text{bulge}}) - S_{\text{gen}}(\gamma_{\text{aptz}})$ is subleading. To see this, note that $\theta_U$ of spheres on the apparent horizon scales linearly with $V$ while $\partial_U S$ is constant. Therefore, if the null shell is sufficiently in the past of $V_0$, the leading generalized entropy increase change between $\gamma_{\text{aptz}}$ and $\gamma_{\text{bulge}}$ is due to bulk entropy.

\section{Comparison with Toy Models} \label{sec:toymodels}
In this section we argue that the size of the Python's Lunch is correctly computing the complexity of decoding the interior modes by explicitly constructing Grover-search-based recovery protocols for simple scrambling quantum circuits. We then explain how this circuit connects to the tensor network explanation of the Python's Lunch geometry given in Sec.~\ref{sec:PLreview}.

We model a black hole formed from collapse and allowed to evolve for time $t$ by a quantum circuit $U$ on $n$ qubits. The starting state is $\ket{0}$, and the circuit has $n t$ two-local gates applied to randomly chosen qubits. We will assume that the time $t$ is much greater than the scrambling time $\log n$ so that the system has time to fully scramble. Finally we extract a small number $h \ll n$ of qubits to be our `Hawking quanta'. For closely related previous constructions, see~\cite{PapRaj13a, PapRaj13b, HarHay13, Yos18, KitYos17, BroGha19}.

We want to understand the complexity of reconstructing the interior partners of these $h$ Hawking quanta for a code space where the Hawking quanta and their partners can be in an arbitrary state (but everything else is fixed). Equivalently, we want to find some unitary circuit $V_{B R_H}$, acting on the remaining $n-h$ qubits $B$ that describe the black hole, together with $h$ reference qubits $R_H$ such that
\begin{align} \label{eq:decoder}
    V_{B R_H} U_{ B H} \ket{0}_{B H} \ket{0}_{R_H}  \approx \ket{\Phi^+}^{\otimes h}_{H R_H} \ket{\psi_0}_B,
\end{align}
where the $h$ Bell pairs $\ket{\Phi^+}$ purify the $h$ Hawking quanta, and $\ket{\psi_0}$ is an arbitrary fixed state. In other words, $V_{B R_H}$ extracts a purification of the Hawking modes $H$ out of $B$ and into $R_H$. Thanks to monogamy of entanglement, the extracted modes must be the interior partners of the Hawking modes.

We assume on general thermalization grounds that the reduced density matrix $\rho_H$ of the $h$ quanta is approximately maximally mixed, or equivalently that a unitary operator $V_{B R_H}$ (simple or otherwise) satisfying \eqref{eq:decoder} exists. This means that the state (which is not normalized) $$2^{h/2} U_{B H} \ket{0}_{B H}$$ can be interpreted as an approximate isometry $V$ from the Hawking quanta $H$ to the black hole Hilbert space $B$. In other words
\begin{align} \label{eq:isometryfact}
    2^h \ket{0}_{B H'} \bra{0}_{B H'} U^\dagger_{B H'} U_{B H} \ket{0}_{B H} \ket{\Phi^+}^{\otimes h}_{H' R_H} \approx \ket{0}_{B H'} \ket{\Phi^+}^{\otimes h}_{H R_H}. 
\end{align}
Here $H'$ is an additional ancilla system, again consisting of $h$ qubits, that makes it easier to describe the desired final circuit. \eqref{eq:isometryfact} is almost exactly the result we want, except that the operator $$2^h \ket{0}_{B H'} \bra{0}_{B H'} U^\dagger_{B H'}$$ is not a unitary (or even an isometry) since it contains the projection operator $\bra{0}_{B H'}$. However, by using Grover search, we can find a unitary circuit $V_{B R_H H'}$ that acts the same way that $$2^h \ket{0}_{B H'} \bra{0}_{B H'} U^\dagger_{B H'}$$ does on the specific state $U_{ B H} \ket{0}_{B H}$. Note that the circuit also acts on the $h$ additional ancilla qubits that we have labeled $H'$ above. However, at the end of the circuit, these ancilla qubits are left unentangled with the rest of the system in the state $\ket{0}$. 
Explicitly, we define the circuit
\begin{align} \label{eq:VBRH}
     V_{B R_H H'}  = \left[\left(\mathds{1} - \ket{\Phi^+}\bra{\Phi^+}^{\otimes h}_{H' R_H} \right)  U_{B H'}\left(\mathds{1} - 2 \ket{0}\bra{0}_{B H'}\right)U^\dagger_{B H'}\right]^{2^{h}\pi/4}.
\end{align}
Then
\begin{align}
    V_{B R_H H'} U_{ B H} \ket{0}_{B H} \ket{\Phi^+}^{\otimes h}_{H' R_H} \approx \ket{0}_B \ket{0}_{H'} \ket{\Phi^+}^{\otimes h}_{H R_H}. 
\end{align}
The argument that the operator $V_{B R_H H'}$ has the desired effect is exactly the same as the one reviewed in Sec.~\ref{sec:PLreview}. In particular, the number of iterations required is proportional to $\sqrt{A} = 2^h$.

We have therefore found that the complexity of reconstructing interior partners is equal to the size of the circuit $n t$ times the number of iterations $O(2^h)$. This agrees with our gravity calculations from Sec.~\ref{sec:PLreview}. 

It also gives an (admittedly somewhat boring) bulk picture of the reconstruction process: at any given stage in the Grover search extraction process, the bulk state is simply a superposition of the initial bulk state and the bulk state with the interior mode successfully extracted. As we repeatedly iterate the process, evolving the bulk backwards and forwards in time, the superposition simply very slowly rotates around, with the amplitude for the extracted state gradually increasing.

\begin{figure}
    \centering
    \includegraphics[width=0.7\textwidth]{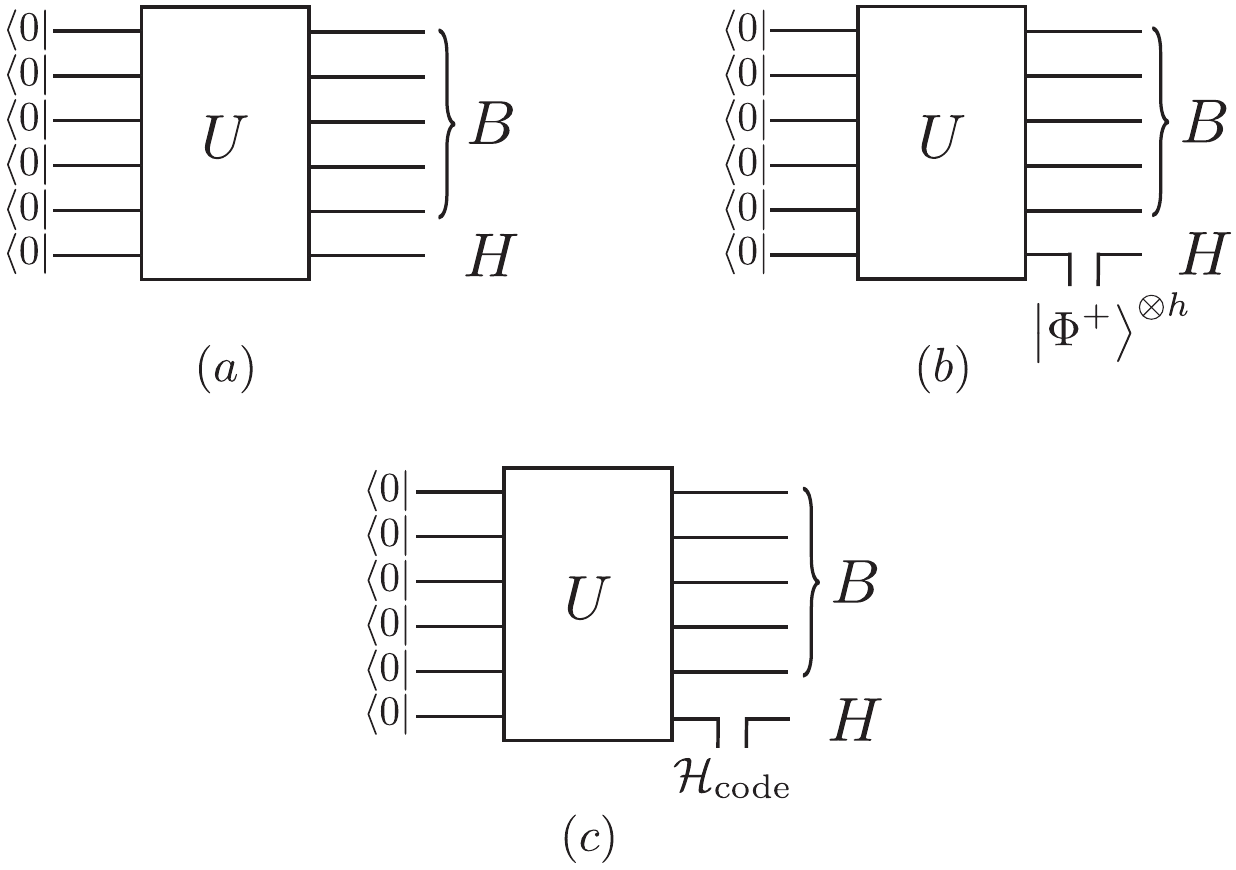}
    \caption{Quantum circuits demonstrating the existence of a lunch associated with the reconstruction of outgoing Rindler-like wave-packets. The circuit $(a)$ can be thought of as time evolution by $U$ on the boundary starting from some initial collapse state, where $H$ represents the state of the $h$ outside Hawking quanta in our code subspace and $B$ represents the remaining black hole. One can equivalently view this circuit as a spatial tensor network representing the interior of the black hole. Then, as shown in $(b)$, the in-plane leg $H$ should be thought of as an entangled state $\ket{\Phi^+}^{\otimes h}$ of the Hawking partners. In $(c)$, the maximally mixed state of the code subspace is represented by the dangling bulk legs which results in a tensor network python's lunch.}
    \label{fig:circuitToTN}
\end{figure}

Let us comment briefly on how the discussion above relates to the more general arguments for the Python's Lunch conjecture based on tensor networks that were given in Sec.~\ref{sec:PLreview}. The first step is to reinterpret the quantum circuit from Fig.~\ref{fig:circuitToTN}(a), as a tensor network describing a spatial slice through the black hole interior rather than a description of boundary time evolution. (The close relationship between exterior time evolution and a spatial slice through the black hole interior is an old story at this point, see e.g.~\cite{HarMal13,StaSus14, Sus14a, SusZha14, Zha20}.) 

In a prima facie paradoxical turn of events, this quantum circuit does not have a Python's Lunch when viewed as a tensor network with all in-plane legs: its cross section is constant in size. However, as shown in Fig.~\ref{fig:circuitToTN}(b), using this as a model for the Hawking quanta and their interior partners requires a reinterpretation of some of the legs of the network as maximally entangled bulk states rather than as in-plane ``area'' legs. In the maximally mixed state for this bulk code subspace, as in Fig.~\ref{fig:circuitToTN}(c), those bulk legs are effectively cut in two, creating a Python's lunch with the same size derived in the gravity calculation. The reconstruction procedure discussed above is thus simply a reinterpretation of the one from Sec.~\ref{sec:PLreview}.

\section{Unknown shocks, BFV states, and Pseudorandomness} \label{sec:BFV}
Pseudorandomness is a fundamental notion in cryptography; it refers to (comparatively) small ensembles of states that are indistinguishable from a completely random state in polynomial time. This notion of polynomial indistinguishability has recently been generalized to quantum states~\cite{JiLiu18}. It was argued by Bouland, Fefferman and Vazirani (BFV) in~\cite{BouFef19} that an easily prepared set of black hole microstates -- specifically black holes formed by collapse together with the application of some randomly chosen shocks -- should form a pseudorandom ensemble. As a consequence, they argue that any boundary reconstruction of the volume of the black hole interior (for a code subspace containing all these states) must be exponentially complex.

If the strong Python's Lunch proposal is true,\footnote{Recall that this is the strictly stronger ``if and only if'' Python's lunch proposal rather than the original version of~\cite{BroGha19}.} then the exponential complexity described by BFV should have a corresponding Python's lunch as its geometric avatar.

Indeed, an equivalent way to say that the BFV ensemble is pseudorandom is to say that the thermal ensemble is a ``coarse-graining'' of the BFV ensemble and so has the same expectation value for all polynomial complexity observables. But, as argued in~\cite{EngWal17a, EngWal17b, BouCha19, EngPen21}, the maximum entropy that such a coarse-grained state can have is given by the generalized entropy of the outermost quantum extremal surface. So, for example, the pseudorandomness of Hawking radiation in an evaporating black hole after the Page time~\cite{KimTan20} follows from the island (the part of the black hole interior in the entanglement wedge of the radiation) being inside of a Python's lunch \cite{BroGha19}. Again, it seems that consistency requires the BFV ensemble to have a Python's lunch -- with an appetizer surface that lies near the event horizon, so that the coarse-grained state is a thermal black hole.

In this section, we show that, while the individual states in the BFV ensemble are simply prepared and therefore have no lunch, the entire ensemble, viewed as a bulk mixed state, \emph{does} have a ``secret'' lunch similar to the ones in Sec.~\ref{sec:maincalc}. This lunch explains the pseudorandomness and exponential reconstruction complexity found by \cite{BouFef19} and thus provides further evidence for the strong Python's Lunch conjecture. 

Let us begin by briefly reviewing the construction of \cite{BouFef19}. We start by preparing a state that admits a simple boundary description: a quantum quench at some fixed time. To be precise, consider an initial vacuum state in a holographic CFT and inject a large amount of energy into the system at $t=0$. As more time goes by the system will thermalize with some large effective temperature $T$ and scramble; in particular, its complexity compared to the $t=0$ state grows linearly with time~\cite{HarMal13, Sus14a, AtiAha17}. The bulk dual of this state is a large AdS black hole formed from collapse of some initial matter at $t=0$, which features a linearly growing maximal volume slice. The complexity=volume proposal of~\cite{StaSus14, Sus14a} accurately reflects this behavior in this system. Suppose now that we are given such a state at some unknown sub-exponential time after $t=0$, and that we want to determine its complexity. From the bulk point of view, the volume of the wormhole is a simple observable (even if it cannot be directly measured by a single bulk observer due to causality constraints). Similarly, on the CFT side, the complexity can be easily determined by seeing how long the state takes to unscramble when evolving the system backwards in time.

\begin{figure}
    \centering
    \includegraphics[width=0.75\textwidth]{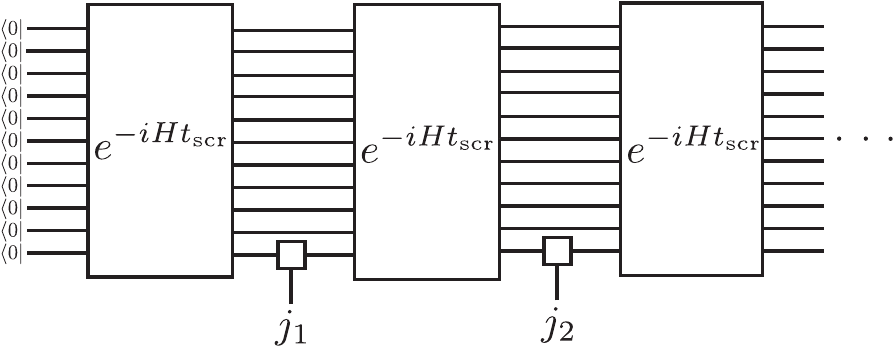}
    \caption{A circuit representation of the BFV ensemble. Every scrambling time, a simple unitary acts on $h$ outside Hawking quanta. This is represented by the small box with label $j_a$ representing the choice of the simple unitary $U_a$, drawn from products of $h$ Pauli operators. After $k$ steps, this results in an ensemble of $4^{kh}$ possible states. Viewing the circuit as a tensor network representing the spatial bulk geometry, the maximally mixed state of the ensemble will be represented by replacing the simple unitaries by two bulk dangling legs, similarly to Fig.~\ref{fig:circuitToTN}$(c)$. This results in multiple python's lunches predicting a complexity of $O(t~2^{kh})$ for distinguishing various states of the ensemble.}
    \label{fig-BFVtensor}
\end{figure}

The pseudorandom set of states constructed by BFV arise when we modify the evolution of the state by inserting $k$ simple boundary unitaries separated by at least a scrambling time, each acting on near boundary thermal modes in equilibrium with the black hole:
\begin{align}
\ket{\psi} = e^{-i H t_{\text{scr}}} U_k e^{-i H t_{\text{scr}}} U_{k-1} \cdots U_1 e^{-i H t_{\text{scr}}} \ket{\psi(t=0)}.
\end{align}
Here $H$ is the boundary Hamiltonian and $t_{\text{scr}}$ denotes the scrambling time. A simple toy model version of this setup, where the black hole is modelled by a large number of qubits, is shown in Fig.~\ref{fig-BFVtensor}.

Even though we define $U_a$ (for $a=1,\cdots, k$) through the bulk, we know they are simple as they act near the boundary. On the gravity side, this can be confirmed since each $U_a$ throws a small amount of energy to the black hole; this has a relatively minor effect on the linear growth of the maximal volume slice. Clearly, if we know the pattern of the unitaries applied, we can repeat the previous procedure to determine the complexity efficiently on the CFT. 

What if we do not know the order and choice of applied unitaries? Specifically, say each $U_a$ is chosen at random from $4$ possible choices. These can be thought of as the four possible Pauli operators (including the identity) that can act on a single qubit. The result is a branching of $4^k$ possible CFT states. On the CFT side, rewinding with $e^{i H t}$ does not reveal the complexity as before since the unitaries prevent the system from unscrambling at $t=0$. On the bulk side, the energies generated by each unitary collide with the initial matter that formed the black hole. The resulting backreaction forms a white hole. 

To unscramble the state, it appears that we need to try acting with all possible combinations of unitaries until we hit upon the right one and the system unscrambles. Naively, this would on average take a time of order $4^k$. By using Grover search, we can obtain a quadratic speed up to $O(2^k)$, but it is not clear how the process can be improved any further.

If it is exponentially complex to unscramble the states, it is presumably also exponentially complex -- from a boundary perspective -- to distinguish the states, or to determine the length of time that the system has been evolving for. On the other hand, in the bulk the states are easily distinguished by looking at the state of the interior. So it appears that the holographic dictionary relating the two must be exponentially complex. While in~\cite{BouFef19} the discussion focused on the complexity of determining the volume of the interior because of the complexity = volume proposal, we emphasize that the same arguments apply to other interior observables such as the state of the shocks, or the state of the initial collapsing matter.

We may now put the strong Python's lunch conjecture to the test: this exponential complexity needs to come from a Python's lunch in the maximally mixed code subspace state -- namely the mixed state describing the full BFV ensemble.

Before directly diving into a gravity calculation, it is helpful to look again at the quantum circuit toy model shown in Fig.~\ref{fig-BFVtensor}. For any fixed $a$, we have a single unitary quantum circuit, i.e. a tensor network with no lunch. However, if we think of the choice of unitary $a$ as a bulk leg that is an unknown input to the tensor network, then suddenly a Python's lunch appears. The size of the lunch (in qubits) is $2k$, which explains the Grover search decoding time of $O(2^k)$ discussed above.

Now let's turn to the gravity calculation. To stay close to the BFV toy model, it is helpful to choose the unitaries $U_a$ so that they act on $h$ fermionic Hawking radiation modes. Choosing the modes to be fermionic allows us to choose the unitaries $U_a$ to genuinely be one of the $4^h$ possible (products of) Pauli operators. We also want to choose the energy $\omega$ of the modes to satisfy $\omega \lesssim T$ so that they are actually in a thermal state as in the toy model.\footnote{If we have $\omega_0 \gg T$ then the unitary $U_a$ can be undone by simply using a channel that maps everything to the ground state (because there is no entanglement with an interior partner that needs to be preserved). Hence the pseudorandomness argument breaks down.}

\begin{figure}
    \centering
    \includegraphics[width=0.75\textwidth]{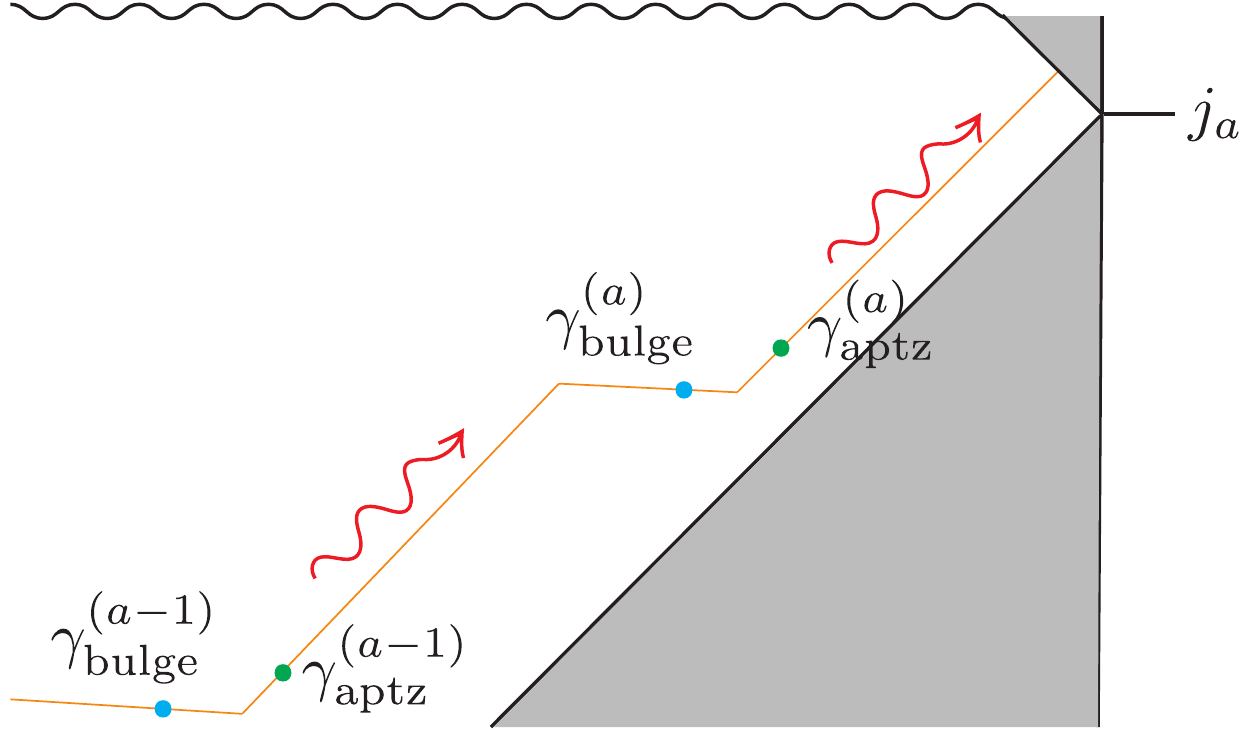}
    \caption{A Wheeler–DeWitt patch in the Penrose diagram of the BFV setup is shown. The mixed state of the ensemble, where we do not know which $U_a$ is applied, results in a state where the Hawking partners are in a maximally mixed state. This results in $\gamma^{(a)}_{\text{aptz}}$ and $\gamma^{(a)}_{\text{bulge}}$ quantum extremal surfaces. Other $U_a$'s not shown in this figure give rise to similar quantum extremal surfaces, e.g. $\gamma^{(a-1)}_{\text{aptz}}$ and $\gamma^{(a-1)}_{\text{bulge}}$, resulting overall in a series of python's lunches.}
    \label{fig-dinner}
\end{figure}

As in Sec.~\ref{sec:maincalc}, acting with a random choice of Pauli $U_a$ disentangles the Hawking modes from their interior partners and thereby nucleates a quantum extremal surface near the horizon.\footnote{We emphasize again that this disentanglement only happens when you look at the mixed state formed by averaging over the random $U_a$. In each individual element of the ensemble, the entanglement remains present and there is no QES.} If we act with multiple unitaries $U_a$, each separated by more than a scrambling time, we actually get a series of multiple lunches, like those discussed in Sec.~\ref{sec:multiple}. These can be characterized by a series of bulges $\gamma^{a}_{\text{bulge}}$ and appetizers $\gamma^{a}_{\text{aptz}}$ as shown in Fig.~\ref{fig-dinner}. The generalized entropy difference between each bulge and its corresponding appetizer follows from the analysis of Sec.~\ref{sec:maincalc}:
\begin{align}\label{eq-deltaSgenbulge}
    S_{\text{gen}}(\gamma^a_{\rm bulge})-S_{\text{gen}}(\gamma^a_{\rm aptz}) \approx 2 h S_{\text{th}}(\omega_0).
\end{align}
\indent On the other hand the difference $S_{\text{gen}}(\gamma^{a}_{\rm bulge})-S_{\text{gen}}(\gamma^{a-1}_{\rm aptz})$ is due to the increase in the area of the black hole caused by the action of $U_{a-1}$. This is equal to $\Delta E / T$ where $\Delta E$ is the increase in the system's energy caused by $U_{a-1}$. If the disentangled modes of $\rho$ had exact Rindler frequencies then we would have $\Delta E=0$ since in that case the state of the exterior modes would be unchanged by applying $U_a$.\footnote{Here we assume that the fermionic modes have energy $\omega_0 \ll T$ and hence are in approximately the maximally mixed state, as in the toy model.} However, disentangling wave-packets results in some energy difference which will in turn increase the area of the black hole: 
\begin{align}
    S_{\text{gen}}(\gamma^{a}_{\rm bulge})-S_{\text{gen}}(\gamma^{a-1}_{\rm aptz}) \sim \frac{h}{(\omega_0 L)^2},
\end{align}
where $L = \log(\delta r_2 / \delta r_1)$ from \eqref{eq-epsilonL1} and  \eqref{eq-epsilonL2} \cite{Marpol15}.\footnote{The fluctuations of $S_{\text{gen}}(\gamma^{a}_{\rm bulge})-S_{\text{gen}}(\gamma^{a-1}_{\rm aptz})$ is of order $\sqrt{h}/(\omega_0 L)$ which is a more important effect than the mean~\cite{Marpol15}. But it can similarly be suppressed compared to the RHS of \eqref{eq-deltaSgenbulge} by taking $L$ to be very large.} Importantly, $L$ can be made arbitrarily large by dialing $\delta r_1$ small. Therefore, from here on we consider a limit where:
\begin{align}
    2 h S_{\text{th}}(\omega_0) \approx S_{\text{gen}}(\gamma^a_{\rm bulge})-S_{\text{gen}}(\gamma^a_{\rm aptz}) \gg S_{\text{gen}}(\gamma^{a}_{\rm bulge})-S_{\text{gen}}(\gamma^{a-1}_{\rm aptz}).
\end{align}
The complexity of reconstructing the interior is then predicted by \eqref{eq:multilunch} to be:
\begin{align}
    C = O\left(N^2 t 2^{k h}\right),
\end{align}
where $N$ is the degree of the boundary gauge group. This complexity is consistent with the analysis of BFV. Moreover it precisely agrees with the complexity predicted for a Grover search reconstruction in the toy model of Fig.~\ref{fig-BFVtensor}.

\section{Discussion} \label{sec:disc}

We have found nontrivial novel QESs that constitute barriers to simple reconstruction of interior Hawking partners even in pure state black holes (that are necessarily not coupled to a reservoir). This derivation immediately dismisses the apparent ``obvious'' contradictions to our strong Python's Lunch proposal that nonminimal QESs are the exclusive source of exponential complexity in the holographic dictionary. We will now discuss some possible generalizations and implications of this result.

\paragraph{Firewalls in typical states:} Possibly the strongest outstanding class of arguments in favor of firewalls -- starting with AMPSS~\cite{AMPSS} and Marolf-Polchinski~\cite{MarPol13} -- are the typicality arguments. These involve reconstructions of interior outgoing modes in a code subspace containing all black hole microstates with smooth horizons, which is generally taken to have dimension of order $e^{A/4G_N}$. It has been pointed out that the paradox presented by these arguments may be resolved via state dependence; specifically a state-dependent boundary representation of the interior Hawking partners (see e.g.~\cite{AMPSS, PapRaj13a, MarPol13}).

One possibility for the origin of such state dependence in bulk reconstruction (even when one has access to the entire asymptotic boundary) is that operations on the interior Hawking modes take you out of the code subspace~\cite{HayPen18, Pen19}. That is, if we would like to reconstruct different states of the outgoing modes, we must increase the code subspace to include various states of the Hawking partners; now the code subspace appears to be of dimension larger than $e^{A/4G_N}$. The maximally mixed state of this larger code subspace leads to a QES on the horizon (to leading order in $G_{N}$), as described in Sec.~\ref{sec:maincalc}.

Importantly, even though this is exactly the same QES that we found in Sec.~\ref{sec:maincalc}, it is now the minimal QES, whereas previously it was simply the appetizer to a Python's Lunch. Reconstructing the interior outgoing modes without state dependence is therefore impossible, rather than simply exponentially complex. This is because, as noted above, in typicality arguments, the dimension of the original code subspace (without the interior outgoing modes) is of order $e^{A/4G_N}$. Since the additional bulk entropy of the outgoing Rindler modes increases the generalized entropy of the empty set but not of the new nontrivial QES, the latter becomes minimal.\footnote{It is important to clarify the novel contribution that we are making here: the idea that matter entropy in the interior of a black hole can exceed the Bekenstein-Hawking entropy goes back to the bag-of-gold geometries considered in \cite{Mar08}; the fact that this could cause the interior of the black hole to be outside the entanglement wedge was discussed already in \cite{HayPen18, Pen19}. However, until Sec. \ref{sec:maincalc} of this paper, the actual QES that is responsible for this latter effect had not been identified.} In contrast, in Sec. \ref{sec:maincalc}, we started with a single black hole microstate. Even after disentangling the Hawking pairs, the empty set therefore still has minimal generalized entropy, and reconstruction is still possible, albeit exponentially complicated.

\paragraph{Other states in $\mathcal{H}_{\text{code}}$:} The state $\rho$ (in \eqref{eq-mixedstate}) is a natural choice for a mixed state in our $\mathcal{H}_{\text{code}}$. However, since bulk reconstruction requires a nontrivial code subspace, and conversely makes sense for any code subspace with more than one state, we expect other states in $\mathcal{H}_{\text{code}}$ to also lead to a nonminimal QES $\gamma_{\text{aptz}}$. Here we will speculate about the type of states in which this happens. For simplicity, let us consider a particular pair of Hawking wave-packet partners across the horizon with $\omega_0 \ll T$. In the HH state, the state of these wave-packets is just the Bell state $\ket{\Phi^+}$. The generalized second law implies that any unitary acting on the outside wave-packet cannot lead to a new quantum extremal surface. Next, let us consider mixed states other than $\rho$. For example, consider the state
\begin{align}
    \tilde{\rho} = \frac{1}{2}(\ket{00}\bra{00}+\ket{11}\bra{11}).
\end{align}
It is easy to see that our Sec. \ref{sec:maincalc} analysis goes through similarly for $\tilde{\rho}$, the only difference being that the RHS of \eqref{eq-dUS} will have an additional factor of $2$ in the denominator which will not change the conclusion that an appetizer surface exists.

What about a product state, e.g. $\ket{00}$? Again, the reduced entanglement across the horizon would allow us to run the argument of Sec. \ref{sec:maincalc} and find an appetizer. In addition, the tensor network analysis of Sec. \ref{sec:toymodels} suggests that in such states a Python's lunch still exists. We can see this by projecting the dangling legs in Fig.~\ref{fig:circuitToTN}(c) to a product state. They would then act as postselected legs in a Python's Lunch tensor network.

These examples suggests that any state in which the mutual information between the wave-packets is reduced compared to $\ket{\Phi^+}$ leads to a Python's Lunch. It should be possible to see this directly from analyzing the quantum expansion, which we leave to future work.

\paragraph{Beyond spherical symmetry:}The HKLL reconstruction~\cite{HamKab05, HamKab06, HamKab06b} of near horizon outgoing Rindler wave-packets fails after a scrambling time due to the transplanckian problem. For s-waves in an Schwarzschild-AdS background, we argued that an appetizer surface $\gamma_{\text{aptz}}$ appears in a mixed state of the corresponding code subspace, explaining the failure of HKLL. However,  since the transplanckian problem also applies to non-spherically symmetric modes in more general backgrounds like Kerr-AdS, our proposal necessarily requires new QESs in mixed states of those code subspaces as well. One way to establish this concretely would be to mirror our spherically symmetric computation of the bulk von Neumann entropy in such mixed states and establish the existence of a quantum anti-normal surface near the horizon.

Here we will briefly discuss an alternative approach to the special case of reconstructing wave-packets localized both in $U$ and the transverse direction (and that thus break the spherical symmetry) in the Schwarzschild-AdS background. We sketch out the direction of a rough plausibility argument directly for the formation of a quantum extremal surface on the horizon. Similarly to the spherically symmetric case, we expect that in the relevant maximally mixed state we have $\Theta_k = O(G_N)$ on the horizon. However, the $\Theta_\ell$ analysis is interestingly different. Here the bulk entropy variation along the $\ell$ direction is only large at some localized portion of the transverse direction. At other transverse directions on the horizon, $\Theta_\ell=0$
requires $\theta_\ell=o(G_N^0)$. It is then necessary for the existence of a quantum extremal surface on the horizon that at leading order we have slices with large $\theta_\ell$ at some transverse location and $\theta_\ell=0$ away from it.

\begin{figure}
    \centering
    \includegraphics[width=0.6\textwidth]{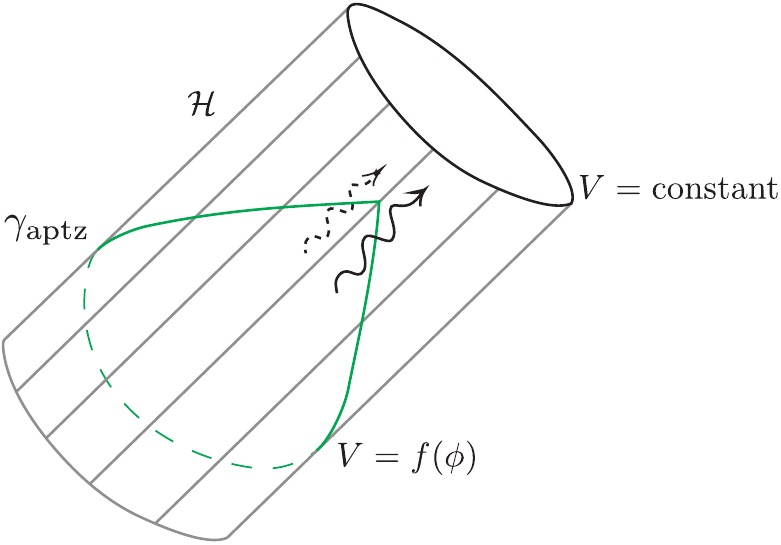}
    \caption{In the maximally (or thermally) mixed state of outgoing Rindler-like wave-packets localized in the transverse direction on the horizon $\mathcal{H}$, we expect that $\gamma_{\text{aptz}}$ would be a non-trivial cut of the horizon like the drop-shaped surface shown (as opposed to a constant $V$ slice). This surface is chosen because it has $\theta_{\ell} \sim O(1)$ around the transverse location of the wave-packets but satisfies $\theta_{\ell}=0$ away from it.}
    \label{fig-drop}
\end{figure}

For concreteness, let us demonstrate the existence of such a slice on a horizon of a non-rotating BTZ black hole with metric:
\begin{align}
ds^2 = \frac{-2 dU dV + R^2 (1-\frac{U V}{2 \ell_{\text{AdS}}^2})^2 d\phi^2}{(1+ \frac{UV}{2\ell_{\text{AdS}}^2})^2},
\end{align}
where $\ell_{\text{AdS}}$ is the AdS radius and $R$ the black hole radius. Now, the slice $V=f(\phi)$ with
\begin{align}
 f(\phi) =f_0 \left(e^\phi + e^{2\pi-\phi}\right),
\end{align}
(where $f_0>0$ is some constant) satisfies:
\begin{align}
\theta_\ell = 2 f_0 (1-e^{2\pi})~\delta(\phi).
\end{align}
Note that this surface has a kink at $\phi=0$ (see Fig.~\ref{fig-drop}) and is smooth away from $\phi=0$. Since well-behaved wave-packets need to be smeared in the transverse direction a bit, we expect that the horizon quantum extremal surface would then need to have a correspondingly smoothed out kink.

In higher dimensions with a non-spherically symmetric background stationary horizon, e.g. AdS-Kerr, and with a more general non-spherically symmetric bulk entropy derivative, finding the QES slice $V = f(y^i)$ (where $y^i$ denotes the transverse direction) would be more complicated. The $\Theta_\ell = 0$ equation is~\cite{EngFis19}
\begin{align}
    \nabla^2 f + 2\chi^i \nabla_i f + (\partial_V \theta_U \rvert_{V=\text{const.}}) f = -\frac{4 G_N}{\sqrt{h(y^i)}}\left.\frac{\delta S_{\text{ren}}}{\delta U(y^i)}\right\rvert_{U=0, V=f(y^i)}
\end{align}
where the LHS is a general expression for $\theta_\ell$ on a stationary horizon involving $\nabla$, the transverse Laplacian and $\chi$ the twist of constant $V$ slices on the horizon, and the RHS is the functional derivative of the renormalized entropy in the $\ell$ direction involving $h(y^i)$, the determinant of the intrinsic metric of the slice. This type of differential equation was discussed in \cite{EngFis19, EngWal18}. Because the marginally trapped surfaces on the horizon are (strictly) stable classically, it is possible to show that the largest real eigenvalue of $L$ is both real and positive~\cite{AndMar05}, so the equation can be inverted. However, depending on the sign of the variation in $S_{\mathrm{gen}}$, it is possible for $f$ to be either positive or negative. For our construction to work in this case, $f$ must be constrained to be strictly non-negative. This is further obfuscated by the fact that $f$ is not differentiable (although this is not fatal, as much of the theory of elliptic operators can be extended to weakly differentiable functions). We leave a further investigation of this to future work.

\subsection*{Acknowledgments}
It is a pleasure to thank Chris Akers, Raphael Bousso, Daniel Harlow, Douglas Stanford, Leonard Susskind, and Edward Witten, for helpful discussions. NE is supported by NSF grant no. PHY-2011905, by the U.S. Department of Energy under grant no. DE-SC0012567 (High Energy Theory research), and by funds from the MIT physics department. GP is supported by the UC Berkeley physics department, the Simons Foundation through the "It from Qubit" program, the Department of Energy via the GeoFlow consortium and also acknowledges support from a J. Robert Oppenheimer Visiting Professorship at the Institute for Advanced Study. ASM is supported by the National Science Foundation under Award Number 2014215.

\appendix

\section{$\partial_U S$ in JT gravity}\label{sec:JTcalc}

In this appendix, we will substantiate our earlier assertion in Sec.~\ref{sec:maincalc} about the inwards variation of the entropy in the HH state. To do so, we will compute $\partial_U S[H(U,V)]$ for conformal matter in the HH state of a finite temperature JT black hole background, and show that it is $O(V)$ near the event horizon at $V=0$.

We work with Kruskal coordinates, where the metric takes the form:
\begin{align}\label{eq-metric2d}
    ds^2 = \frac{-4}{(1+UV)^2} dU dV.
\end{align} 
We have set the AdS scale to zero and as usual in Kruskal coordinates, the horizons are at $UV=0$ and the ``singularity (where the dilaton diverges to $-\infty$) is at $UV=1$. The dilaton profile $\phi =\phi_0 + \phi$ is
\begin{align}
    \phi = 2 \pi T \bar{\phi}_r \frac{1-UV}{1+UV}.
\end{align}

We work with the HH state for the conformal matter in this geometry, which can be obtained from the Minkowski vacuum via a Weyl rescaling: 
\begin{align}\label{eq-2dweyl}
    \Omega(U,V) = \frac{\sqrt{2}}{1+UV}.
\end{align}
Now, consider the von Neumann entropy of a region with $\Delta U$ and $\Delta V$ extents in 2D Minkowski vacuum~\cite{CalCar04}
\begin{align}\label{eq-2dMinkS}
    S_{\text{flat vacuum}} = \frac{c}{6} \log \left( \frac{\Delta U}{\sqrt{\epsilon^{U}_1 \epsilon^{U}_2}}\right) + \frac{c}{6} \log \left( \frac{\Delta V}{\sqrt{\epsilon^{V}_1 \epsilon^{V}_2}}\right),
\end{align}
where $\epsilon^{U}_{1}$ and $\epsilon^{V}_{1,2}$ are respectively the $U$ and $V$ cutoff at endpoints 1 and 2 of the interval. We can compute the von Neumann entropy on a region in the black hole spacetime by rescaling $\epsilon^{U}_{1}$ and $\epsilon^{V}_{1,2}$ in Eq. \eqref{eq-2dMinkS} appropriately with the Weyl factor \eqref{eq-2dweyl}. There is a subtlety, however, since the black hole spacetime has reflecting boundary conditions at $UV=-1$ and the Minkowski vacuum does not.  we can still compute $\partial_U S[H(U,V)]$ in the black hole spacetime from the entropy of null lines that extend from some point in $(U,V)$ to the $UV=-1$ line along the $U$ direction in Minkowski and compute the $U$ derivative. This is because the reflecting boundary conditions convert outgoing modes to infalling modes in AdS, but the entropy variation on a null (along $U$) interval is not affected by this since ingoing modes do not register on it. The null line has $\Delta U = U + 1/V$. By incorporating the warped factor into $\epsilon^U$ of Eq.\eqref{eq-2dMinkS}, we obtain:
\begin{align}
    \partial_U S[H(U,V)] = \partial_U \frac{c}{6} \log \left( \frac{U+1/V}{\sqrt{\epsilon^{U} \frac{\sqrt{2}}{(1+U V)}}}\right) = \frac{c}{8} \frac{V}{1+ UV}.
\end{align}
In the region near the future horizon, i.e. $V>0$ and $U V \ll 1$, we have $\partial_U S=\mathcal{O}(V)$. This is always subleading by a factor of $G_{N}$ compared to $(1/4G_N) \partial_U \phi$. A consequence of this is that the only quantum extremal surface in the HH state is the bifurcation surface.

\bibliographystyle{jhep}
\bibliography{all}

\end{document}